\documentclass[aps,prx,amsmath,amsfonts,superscriptaddress,twocolumn,preprintnumbers,floatfix,longbibliography]{revtex4-1}
\usepackage{amssymb}
\usepackage{graphicx}
\usepackage{appendix}
\usepackage{color}
\usepackage{subfigure}
\usepackage[colorlinks=true,linkcolor=blue,filecolor=blue, urlcolor=blue,citecolor=blue]{hyperref}

\begin{document}

\title{Orbitally-selective Breakdown of the Fermi Liquid and Simultaneous Enhancement of Metallic and Insulating States in Correlated Multi-band Systems with Spin-orbit Coupling}

\author{Ze-Yi Song}
\affiliation{Shanghai Key Laboratory of Special Artificial Microstructure Materials and Technology, School of Physics Science and engineering, Tongji University, Shanghai 200092, P.R. China}
\author{Xiu-Cai Jiang}
\affiliation{Shanghai Key Laboratory of Special Artificial Microstructure Materials and Technology, School of Physics Science and engineering, Tongji University, Shanghai 200092, P.R. China}
\author{Yu-Zhong Zhang}
\email[Corresponding author.]{Email: yzzhang@tongji.edu.cn}
\affiliation{Shanghai Key Laboratory of Special Artificial Microstructure Materials and Technology, School of Physics Science and engineering, Tongji University, Shanghai 200092, P.R. China}

\date{\today}

\begin{abstract}
We show that spin-orbit coupling (SOC) plays Janus-faced roles on the orbitally-selective Mott transitions in a three-orbital Hubbard model with crystal field splitting at a specific filling of $2/3$, which is a minimal Hamiltonian for ruthenates. While the SOC favors metallic state due to enhancement of orbital hybridization at smaller on-site Coulomb repulsions, it stabilizes the Mott insulating state ascribed to lifting of orbital degeneracies and enhancement of band polarizations at larger electronic interaction. Moreover, an orbitally-selective non-Fermi liquid (OSnFL), where breakdown and retention of the Fermi liquid coexist in different orbitals, emerges between the orbitally-selective Mott phase and the Fermi-liquid state. This novel state can be used to account for the exotic metallic behavior observed in 4$d$ materials, such as Ca$_{1.8}$Sr$_{0.2}$RuO$_4$, Ba$_2$RuO$_4$ under strain and Sr$_2$RuO$_4$ under uniaxial pressure. We propose that orbitally-selective Kondo breakdown may account for the OSnFL.

\end{abstract}

\maketitle

\section{introduction}
Multi-orbital correlated electronic systems have been extensively investigated for decades since the discoveries of iron-based superconductors~\cite{KamiharaJACS2006,KamiharaJACS2008}, and 4$d$, 5$d$ materials like ruthenates and iridates~\cite{WitczakARCMP2014,RauARCMP2016,MartinsJPCM2017}, etc. Interplay of the following factors like the kinetic energy, crystal field splitting, spin-orbit coupling (SOC), Hund's rule coupling and the Hubbard interaction dominates various exotic properties of these systems. For example, sizable SOC and Coulomb repulsion in 4$d$ and 5$d$ materials~\cite{MartinsJPCM2017} lead to unconventional superconductivity~\cite{MengPRL2014,ChaloupkaPRL2016}, SOC-assisted Mott transition~\cite{KimScience2009,ChaloupkaPRL2010,CominPRL2012,PlumbPRB2014,WinterPRB2016,LangPRB2016}, quantum spin liquid~\cite{ZhouRMP2017}, spin-orbit exciton condensation~\cite{KhaliullinPRL2013,SatoPRB2019}, and exotic magnetic order~\cite{KhaliullinPRL2013,ChenPRB2010,ChenPRB2011,MeeteiPRB2015}. Among all the phenomena mentioned above, SOC-assisted Mott transition is of particular interest, where the SOC lifts orbital degeneracy, resulting in an effective half-filled system. Then, the Mott transition occurs at intermediate Coulomb interaction, as observed in Sr$_2$IrO$_4$~\cite{KimScience2009}.

Meanwhile, the Hund's rule coupling served as a band decoupler~\cite{MediciPRB2011}, together with the crystal field splitting which lowers the orbital degeneracy, open a promising way to generate orbital selectivity which is believed to widely exist in ruthenates~\cite{GeorgesARCMP2013,GorelovPRL2010,MravljePRL2011,StrickerPRL2014,DangPRL2015,DangPRB2015,SutterPRB2019,KuglerPRL2020} and iron-based superconductors~\cite{GeorgesARCMP2013,HauleNJP2009,YinNatPhys2011,NicolaPRB2013,YiPRL2013,SprauScience2017}. The corresponding metallic state, so-called Hund's metal, with electronic correlations dominated by Hund's coupling, rather than the Hubbard interaction, are of current intensive interest since it may be responsible for exotic metallic behavior~\cite{GeorgesARCMP2013} and unconventional superconductivity~\cite{HauleNJP2009}. And under certain circumstances~\cite{KogaPRL2004,WernerPRL2007,MediciPRL2009,LeePRB2011,SongNJP2015}, the orbitally-selective Mott (OSM) transition~\cite{AnisimovEPJB2002} takes place, where partial bands become Mott-insulating and the others remain metallic.

However, comprehensive understanding of quantum phase transitions in real materials with multiple active orbitals is still missing~\cite{AnisimovEPJB2002,ShimoyamadaPRL2009,NeupanePRL2009,GorelovPRL2010,SutterPRB2019} since the SOC and the crystal field splitting are always separately taken into account~\cite{HuangPRB2012,DuEPJB2013,KimPRL2017,TrieblPRB2018,PiefkePRB2018} when the electronic correlations are theoretically treated within some reliable approximations like the dynamical mean field theory (DMFT)~\cite{GeorgesRMP1996}. Such an unrealistic modelling may result in misleading of the effect of SOC on the Hund's metals, the orbital selectivity, and the Mott insulators.

\begin{figure}[htbp]
	\includegraphics[width=0.48\textwidth]{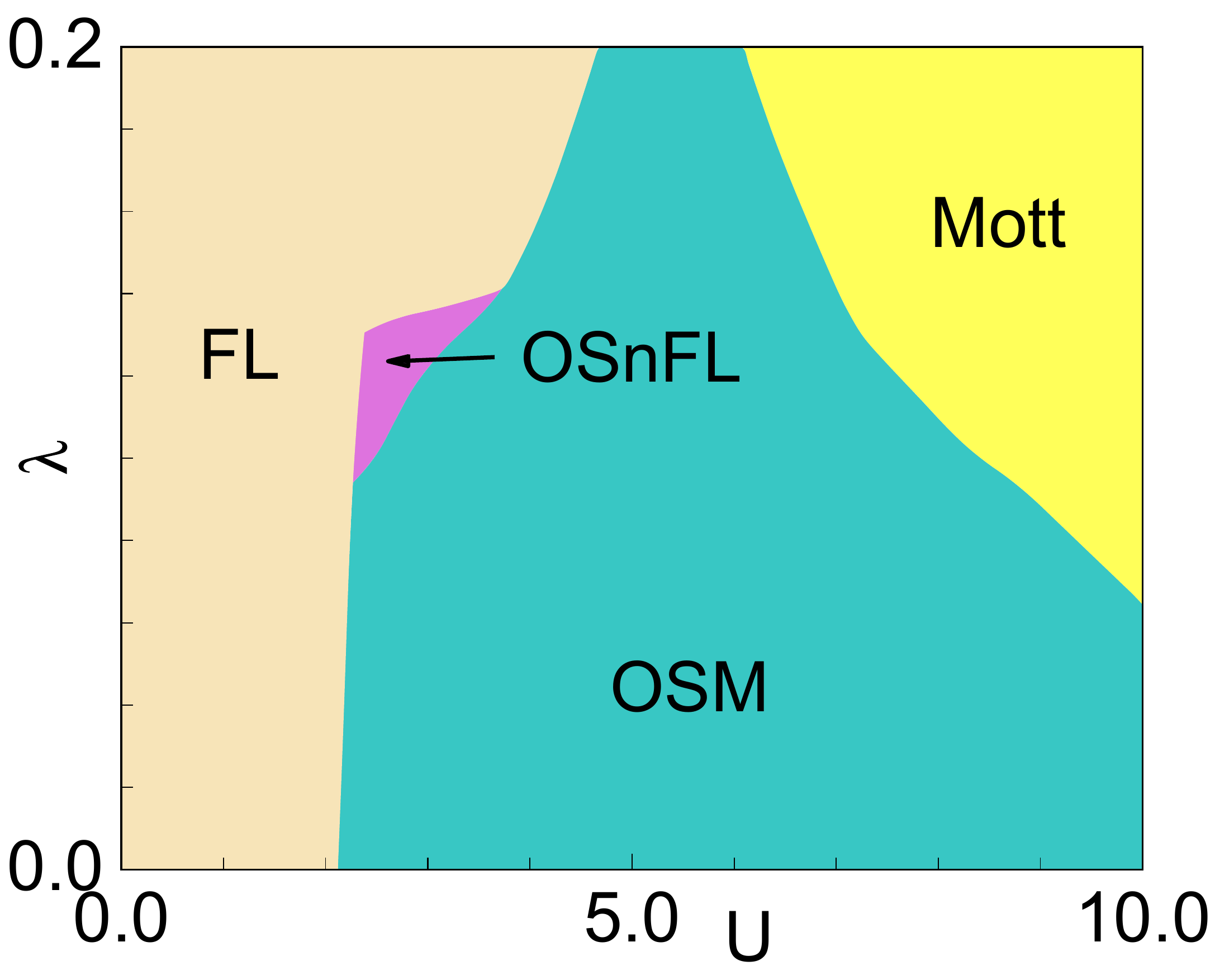}
	\caption{(color online) The schematic phase diagram for the three-orbital Hubbard model in the plane of $\lambda$ and $U$.  FL, OSM, OSnFL, and Mott denote the Fermi liquid,  orbitally-selective Mott phase, orbitally-selective non-Fermi liquid, and Mott insulator, respectively.}
	\label{phase_diagram}
\end{figure}

In this article, we have studied a three-orbital Hubbard model to investigate the influence of SOC on the orbital-differentiated correlations induced by both the lifted degeneracies due to the crystal field splitting and the orbital decoupling ascribed to the Hund's coupling at zero temperature. The main results are schematically summarized in Fig.~\ref{phase_diagram}. In the absence of SOC, two phases are identified within the interacting range we studied, including the Fermi-liquid state and the OSM state. The SOC is found to effectively suppress the electronic correlations at small Coulomb interaction and favor the Fermi-liquid state. The competition between SOC and electronic correlations leads to the emergence of an exotic metallic state~\cite{CooperScience2009}, characterized by the coexistence of the Fermi-liquid state and non-Fermi-liquid state in different orbitals, which can account for orbitally-selective breakdown of the Fermi liquid observed in Ca$_{1.8}$Sr$_{0.2}$RuO$_4$ by recent angle-resolved photoemission spectroscopy (ARPES) experiments~\cite{SutterPRB2019} and early transport measurements~\cite{NakatsujiPRL2003}. Contrastly, the SOC enhances electronic correlations at large Coulomb repulsion and results in reduced critical values of phase transitions from the OSM state to the Mott insulator.

\section{Model and Method}

To obtain above phase diagram, the three-orbital Hubbard model on the bethe lattice is considered, which is defined as
\begin{align}\label{Hubbardmodel}
H&=t\sum_{\langle{ij}\rangle\alpha\sigma}C^{\dag}_{i\alpha\sigma}C_{j\alpha\sigma}+\sum_{i\alpha\sigma}(\Delta_{\alpha}-\mu)n_{i\alpha\sigma}	\nonumber\\
	&+U\sum_{i\alpha}n_{i\alpha\uparrow}n_{i\alpha\downarrow}+(U^{\prime}-J_z)\sum_{i\alpha>\beta\sigma}n_{i\alpha\sigma}n_{i\beta\sigma}\nonumber\\
	&+U^{\prime}\sum_{i\alpha>\beta\sigma}n_{i\alpha\sigma}n_{i\beta\bar\sigma}-J_{f}\sum_{i\alpha>\beta}\left[S^+_{i\alpha}S^-_{i\beta}+S^-_{i\alpha}S^+_{i\beta}\right]\nonumber\\
	&+J_p\sum_{i\alpha\neq\beta}C^{\dag}_{i\alpha\uparrow}C^{\dag}_{i\alpha\downarrow}C_{i\beta\downarrow}C_{i\beta\uparrow} + H_{\text{SOC}},
\end{align}	
where $t$ denotes the nearest-neighbor hopping, $\Delta_{\alpha}$ represents crystal field splitting for $t_{2g}$ basis with orbital indices $\alpha=\{yz,xz,xy\}$, and $\mu$ is the chemical potential. $U$ and $U^{\prime}$ are the onsite intraorbital and interorbital Coulomb repulsions, respectively. The Hund's rule couplings consist of the Ising-type coupling $J_z$, the spin-flip term $J_f$ and pair-hopping term $J_p$. The relationship $U=U^{\prime}+2J_z$ is employed to assure the electronic interaction rotationally invariant. $C^{\dag}_{i\alpha\sigma} (C_{i\alpha\sigma})$ creates (annihilates) an electron with spin $\sigma$ in orbital $\alpha$ of lattice site i. $S$ and $n$ represent the spin and particle number operators, respectively. The relativistic SOC reads
\begin{align}\label{HSOC}
H_{\text{SOC}}=\lambda\sum_{i\alpha\beta}\sum_{\sigma_1\sigma_2}\langle\alpha|{\mathbf L}_i|\beta\rangle\langle\sigma_1|{\bf{}S}_i|\sigma_2\rangle{}C^{\dag}_{i\alpha\sigma_1}C_{i\beta\sigma_2},
\end{align}
where $\lambda$ is the strength of SOC, $\bf L$ is the local orbital angular momentum operator. The matrix representations of $L=2$ in the $t_{2g}$ basis are the same as the ones for $L=1$ in cubic basis except for a sign in accordance with the T-P correspondence~\cite{MartinsJPCM2017,Sugano1970}.

We employed the DMFT in combination with exact diagonalization (ED)~\cite{GeorgesRMP1996} as impurity solver to solve the model (\ref{Hubbardmodel}) on the bethe lattice with infinite coordinates at a filling of $n=2/3$, namely $4$ electrons in $3$ orbitals, and at zero temperature. The noninteracting density of states is $\rho_{\alpha}(\omega)=\frac{2}{\pi{D}^2}\sqrt{D^2-\omega^2}$ and the half bandwidth $D$ is used as the energy unit. The effective inverse temperature was set to $\beta{D}=200$ which serves as a low-frequency cutoff. On the bethe lattice, the DMFT self-consistent conditions simply read $\hat\Delta(\omega)=\frac{D^2}{4}\hat{G}(\omega)$, where $\hat\Delta(\omega)$ is a matrix for hybrid functions, $\hat{G}(\omega)$ for local lattice Green's functions. Totally 6 bathes were used to fit the hybrid function $\hat\Delta(\omega)$. In our calculations, the orbital degeneracy is lifted by crystal field splitting (i.e., $\Delta_{yz}=\Delta_{xz}\neq\Delta_{xy}$), leading to a nondegenerate $d_{xy}$ orbital and doubly degenerate $d_{yz/xz}$ orbitals. We fix electronic populations to be $(1.5,1.5,1.0)$ in accordance with that of Ca$_{1.8}$Sr$_{0.2}$RuO$_4$~\cite{SutterPRB2019} in the absence of SOC, which can be realized by tuning the orbitally-dependent potential $\Delta_{\alpha}$~\cite{MediciPRL2009}. The calculations were performed in the paramagnetic state with isotropic Hund's coupling where $J_z=J_f=J_p$. The broadening factor $\eta=0.02D$ is used to calculate real-frequency dynamical quantities including the Green's functions, self-energies, and dynamical susceptibilities for spin, orbital, and total angular momentums.

\section{results}

\begin{figure}[htbp]
	\includegraphics[width=0.5\textwidth,height=0.16\textheight]{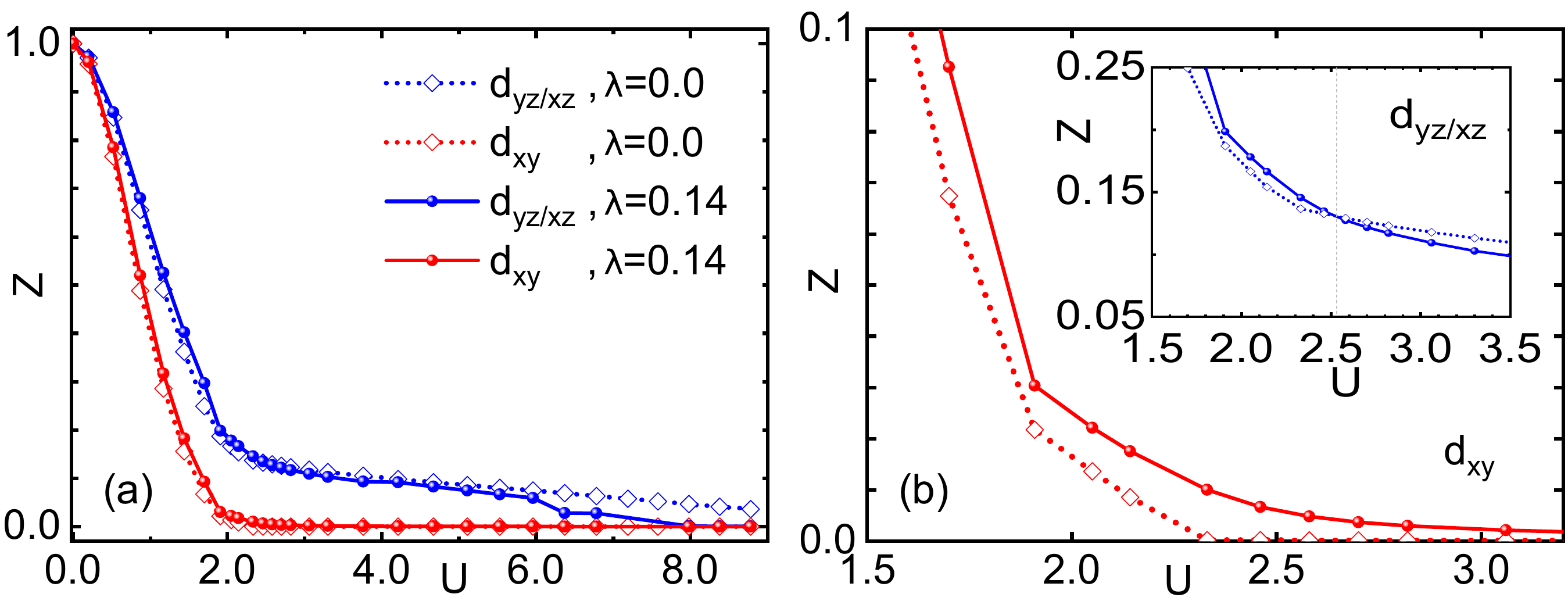}
	\caption{(color online) (a) The quasiparticle residue $Z$ of $t_{2g}$ orbitals as a function of $U$ at $\lambda=0.0$ and $\lambda=0.14$, (b) a blowup for $Z$ of $d_{xy}$ orbital near the critical point $U^c_{xy}(\lambda=0.0)$ of about 2.3, those for $d_{yz/xz}$ orbitals are shown in the inset.  }
	\label{residuewZ}
\end{figure}

In order to explore the influence of SOC on the electronic correlations in multi-band systems, we have calculated the quasiparticle residue
$Z_{\alpha}=(1-\frac{\partial{\text{Re}}\Sigma_{\alpha}(\omega)}{\partial\omega}|_{\omega\to0})^{-1}$, as shown in Fig.~\ref{residuewZ} (a), where ${\text{Re}}\Sigma_{\alpha}(\omega)$ is the real part of self-energies of $t_{2g}$ orbitals. In the absence of SOC, $Z_{xy}$ is rapidly suppressed as $U$ increases and vanishes at a critical value of $U^c_{xy}(\lambda=0.0)\approx2.3$, indicating a Mott-Hubbard gap opens in $d_{xy}$ band. In contrast, although $Z_{yz/xz}$ is drastically reduced at the beginning, it remains finite in a wide region of $U$ with $U^c_{yz/xz}(\lambda=0.0)$ much larger than $U^c_{xy}(\lambda=0.0)$, which suggests an occurrence of the OSM phase where electrons in $d_{xy}$ band become Mott-localised and those in $d_{yz/xz}$ bands remain itinerant. This is in excellent agreement with previous results~\cite{MediciPRL2009}.

When the SOC is taken into account, e.g. $\lambda=0.14$, the quasiparticle residues behave distinctly at small and large Coulomb repulsions in comparison to those at $\lambda=0.0$. At large $U$, the SOC enhances the electronic correlations in $d_{yz/xz}$ bands as indicated by suppression of $Z_{yz/xz}$, leading to a smaller critical value of Mott transition $U^c_{yz/xz}(\lambda=0.14)$ than that of $\lambda=0.0$ case. Since the Mott-insulating state with a vanishing $Z_{xy}$ preserves in $d_{xy}$ band, it suggests that the SOC cooperates with the Coulomb interaction and stablizes the Mott-insulating ground state. Conversely, at small $U$, the SOC suppresses the electronic correlations in all bands as inferred by larger values of $Z$ in comparison to those of $\lambda=0$ case as shown in Fig.~\ref{residuewZ} (b) which is a blowup of Fig.~\ref{residuewZ} (a) at small $U$ region. This gives rise to an intersection of $Z_{yz/xz}$ between $\lambda=0.14$ and $\lambda=0.0$ cases at $U\approx2.5$ as seen in the inset of Fig.~\ref{residuewZ} (b). Furthermore, the quasiparticle residue $Z_{xy}$ of $\lambda=0.14$ case remains nonzero for a wide range of $U>U^c_{xy}(\lambda=0.0)$, suggesting that a transition from  Mott insulator to metal may take place in $d_{xy}$ band when SOC is turned on. This indicates that the SOC competes with the on-site Coulomb repulsion and favors a metallic ground state at small $U$. Obviously, the SOC shows opposite effects on the Mott transitions in multi-band systems.

\begin{figure}[htbp]
	\includegraphics[width=0.48\textwidth,height=0.28\textheight]{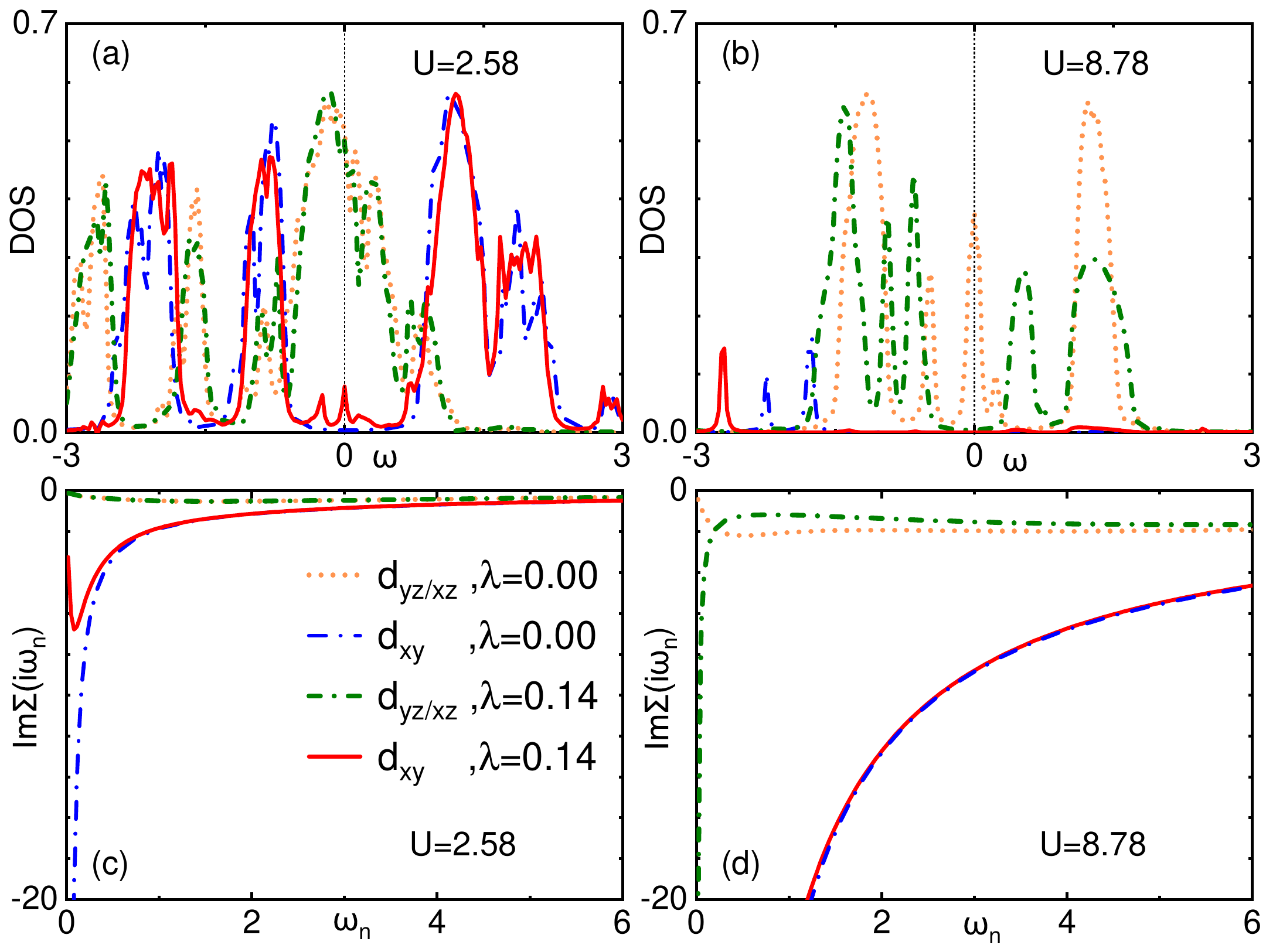}
	\caption{(Color online) (a-b) Density of states projected onto the $t_{2g}$ orbitals and (c-d) the imaginary part of self-energy on the Matsubara axis at $U=2.58$ and $U=8.78$ for $\lambda=0.0$ and $\lambda=0.14$.}
	\label{DOS_SIGMA_T2G}
\end{figure}

We further demonstrate the pronounced effects of the SOC in the vicinity of phase transitions in Fig.~\ref{DOS_SIGMA_T2G} where the calculated density of states projected onto the $t_{2g}$ basis and corresponding imaginary part of the Matsubara self-energy Im$\Sigma_{\alpha}(i\omega_n)$ are exhibited. In the absence of the SOC, at both $U=2.58$ and $U=8.78$, the density of states at the Fermi level vanishes in $d_{xy}$ orbital and Im$\Sigma_{xy}(i\omega_n)$ diverges in proximity to zero frequency, which is a typical character of Mott insulator. Meanwhile, the $d_{yz/xz}$ orbitals behave as Fermi liquids since the density of states is finite at $\omega=0$ and Im$\Sigma_{yz/xz}(i\omega_n)$ approches zero at low frequency. It suggests that the system is in the OSM phase. When $\lambda$ increases to 0.14, the SOC exhibits distinct effects at small and large $U$, respectively. At $U=8.78$, while the SOC barely influences the nature of $d_{xy}$ orbital, the complete suppression of density of states of $d_{yz/xz}$ orbitals at the Fermi level and the divergence of Im$\Sigma_{yz/xz}(i\omega_n)$ as $\omega_n$ goes to zero, suggests an appearance of the SOC-assisted Mott phase. On the other hand, at $U=2.58$, the SOC drastically affects the properties of $d_{xy}$ orbital but hardly influences that of $d_{yz/xz}$ orbitals. The appearance of central peak of density of states for $d_{xy}$ orbital at the Fermi level and Im$\Sigma_{xy}(i\omega_n)$ extrapolating to zero as $\omega_n \rightarrow 0$ suggests a SOC-induced Fermi liquid behavior.

The occurrence of SOC-assisted Mott phase at large $U$ can be easily understood within $|J,\pm{m}\rangle$ basis where the local Hamiltonian~(\ref{HSOC}) can be diagonalized. Here $J$ denotes the total angular momentum and $\pm{m}$ represents its projection in $z$ direction. It is found that the SOC enhances band polarizations and leads to a full occupation of the $|\frac{3}{2},\pm\frac{3}{2}\rangle$ bands at $\lambda_c$ of about $0.1$. Then, the rest two electrons reside in $|\frac{3}{2},\pm\frac{1}{2}\rangle$ and $|\frac{1}{2},\pm\frac{1}{2}\rangle$ bands, resulting in an effective half-filled system, rather than originally $4$ electrons per $3$ orbitals. Therefore, the SOC favors Mott transition at large $U$ since the effective filling is changed.

On the contrary, in the small $U$ region, an OSM phase requires decoupling between Mott-insulating $d_{xy}$ orbital and metallic $d_{yz/xz}$ orbitals, which is originally fulfilled by Hund's rule interaction in the absence of SOC. However, the SOC introduces coupling between the $d_{xy}$ and $d_{yz/xz}$ orbitals, leading to the enhancements of both orbital fluctuations and the kinetic energies of all orbitals. Therefore, the SOC tends to destroy the Mott phase of $d_{xy}$ orbital and favors metallic state due to the increase of bandwidth and decrease of band decoupling at small $U$.

\begin{figure}[htbp]
	\includegraphics[width=0.48\textwidth]{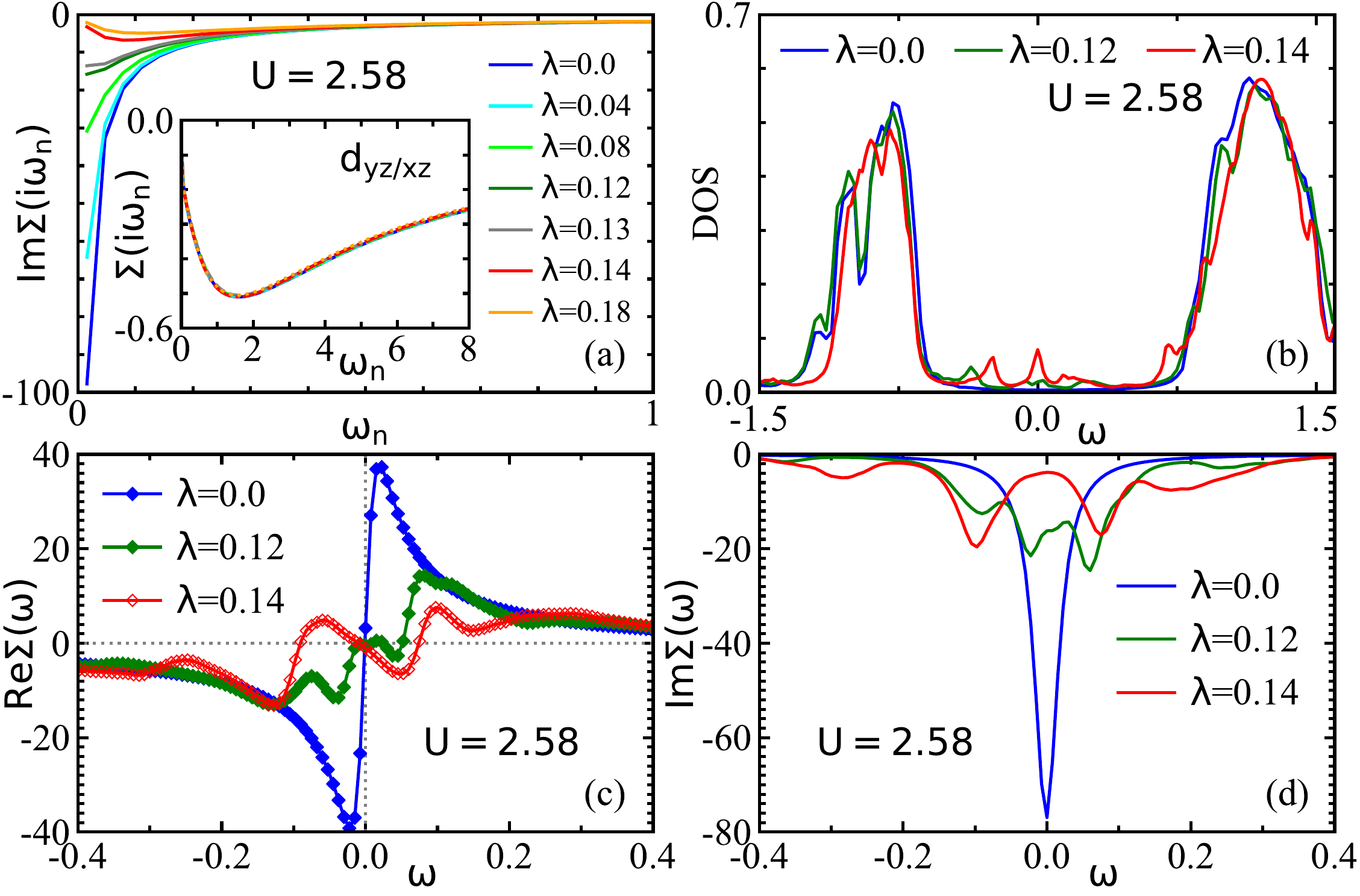}
	\caption{(color online) The influence of spin-orbit coupling on the imaginary part of Matsubara self-energy for $d_{xy}$ band, those for $d_{yz/xz}$ bands are shown in the inset. (b) Density of states for $d_{xy}$ band in the Mott ($\lambda=0.0$), non-Fermi-liquid ($\lambda=0.12$) and the Fermi-liquid ($\lambda=0.14$) states. (c) Real and (d) imaginary part of self-energy of $d_{xy}$ band on the real frequency axis. Here, the electron repulsion is fixed at $U=2.58$.}
	\label{NFL_Sigma_DOS}
\end{figure}

Besides the opposite effect of SOC on electronic correlations, it is also interesting to find an intermediate phase, called orbitally-selective non-Fermi liquid (OSnFL) where breakdown of the Fermi liquid happens only in $d_{xy}$ orbitals, emerging between the Fermi-liquid state and the OSM phase in the small $U$ and intermediate $\lambda$ region, as presented in Fig.~\ref{phase_diagram}. This exotic metallic state can be clearly identified by the imaginary part of the Matsubara self-energy Im$\Sigma_{\alpha}(i\omega_n)$ as shown in Fig.~\ref{NFL_Sigma_DOS} (a). For example, at $U=2.58$, while Im$\Sigma_{yz/xz}(i\omega_n)$ always goes to zero as $\omega_n \rightarrow 0$, indicating the Fermi liquids in $d_{yz/xz}$ bands, (see the inset of Fig.~\ref{NFL_Sigma_DOS} (a)), Im$\Sigma_{xy}(i\omega_n)$ extrapolates to a finite value as $\omega_n \rightarrow 0$ around $\lambda \approx 0.12$, which is in sharp contrast to the divergent behavior for $\lambda \le 0.08$ and the tendency to approach zero for $\lambda \ge 0.14$. The finite scattering rate suggests a finite lifetime of quasiparticles and a breakdown of the Fermi liquid at zero temperature.

The self-energy $\Sigma_{xy}(\omega)$ on the real frequency axis again reveals the SOC-induced non-Fermi-liquid behavior in $d_{xy}$ band. Fig.~\ref{NFL_Sigma_DOS} (c) and (d) shows the real part of self-energy Re$\Sigma_{xy}(\omega)$ and imaginary part of self-energy Im$\Sigma_{xy}(\omega)$ at $U=2.58$ for $\lambda=0.0$, $\lambda=0.12$ and $\lambda=0.14$. When $\lambda=0.12$, the positive slop of Re$\Sigma_{xy}(\omega)$ at the Fermi level and the finite Im$\Sigma_{xy}(\omega)$ suggests the breakdown of quasiparticle picture. The development of additional low-energy poles of $\omega+\mu-\varepsilon-\text{Re}\Sigma_{xy}(\omega)=0$ close to the Fermi level and the finite scattering rate at $\omega=0$ lead to the appearance of a pseudogap in $d_{xy}$ band, as depicted in Fig.~\ref{NFL_Sigma_DOS} (b), reminiscent of that observed in Hubbard model within cluster DMFT~\cite{ZhangPRB2007}. This is strikingly different from those for $\lambda=0.0$ and $0.14$. For the former, the divergent Im$\Sigma_{xy}(\omega)$ at $\omega=0$ and the steep positive slop of Re$\Sigma_{xy}(\omega)$ suggest a Mott-insulating state. In contrast, for the latter, the Im$\Sigma_{xy}(\omega)$ can be fitted by $\omega^2$ and the Re$\Sigma_{xy}(\omega)$ is linearly $\omega$-dependent, indicating a typical Fermi-liquid behavior at zero temperature. Fig.~\ref{NFL_Sigma_DOS} (b) shows the density of states for $d_{xy}$ band in the Mott, pseudogap, and the Fermi-liquid states at $\lambda=0.00$, $0.12$ and $0.14$, respectively.

\begin{figure}[thbp]
	\includegraphics[width=0.48\textwidth]{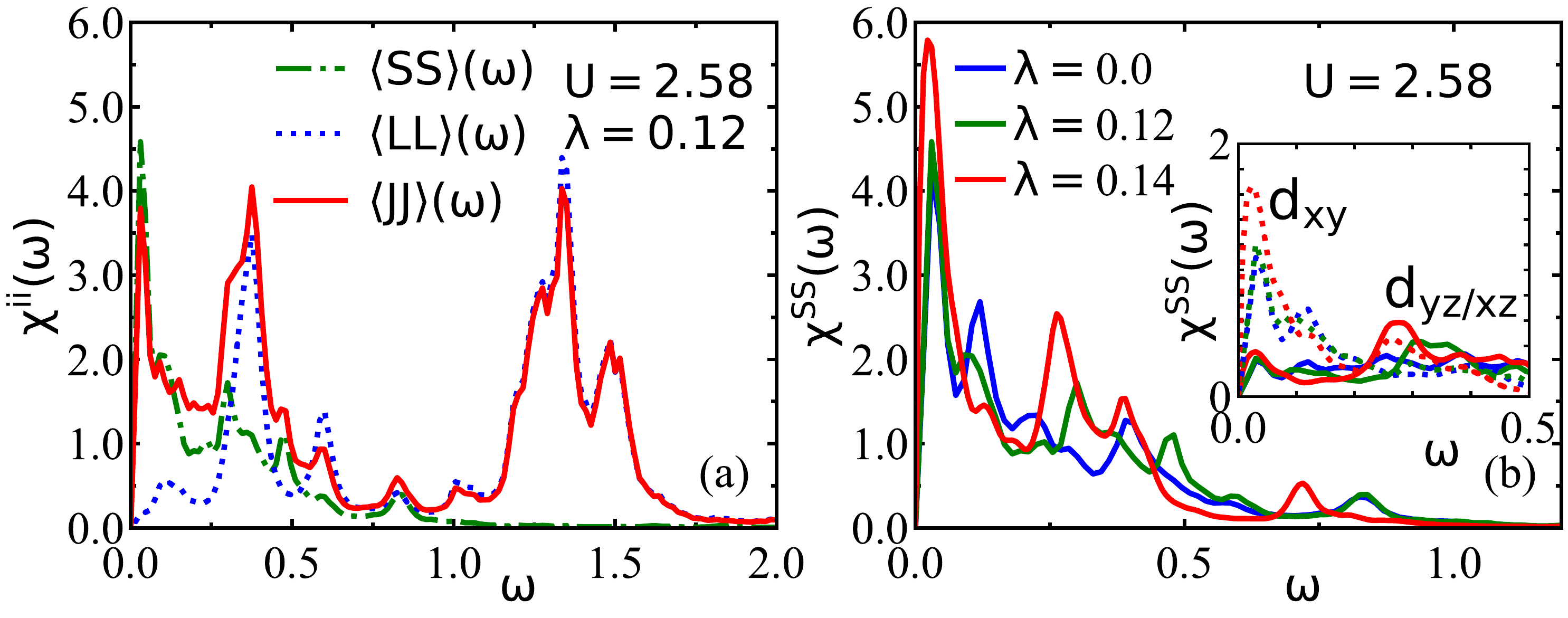}
	\caption{(color online) (a) Dynamical susceptibilities for spin ($\bf S$), orbital ($\bf L$) and total ($\bf J$) angular momentums at $U=2.58$ and $\lambda=0.12$. (b) Dynamical spin susceptibilities at $U=2.58$ for $\lambda=0.0$, $\lambda=0.12$ and $\lambda=0.14$ cases, the inset shows corresponding orbitally-resolved dynamical spin susceptibilities, where solid lines denote $d_{yz/xz}$ orbitals and dashed lines $d_{xy}$ orbital.}
	\label{susceptibility}
\end{figure}

Now we discuss the underlying physics for the OSnFL state. Since the nature of correlated metals is controlled by the low-energy excitations in the absence of SOC for multi-band systems~\cite{SongPRB2017,KuglerPRB2019}, we performed similar calculations for the dynamical susceptibilities $\chi^{ii}(\omega)$ of total, orbital, and spin angular momentums with $i=J$, $L$, and $S$, respectively, at $U=2.58$ and $\lambda=0.12$, as shown in Fig.~\ref{susceptibility} (a). The dynamical spin susceptibilities are defined as
\begin{align}\label{SS_correlation}
	\chi^{SS}(t)=-i\theta(t)\langle|[S(t),S]_{-}|\rangle,
\end{align}
where $\theta(t)$ is a step function, $|\rangle$ is the ground state and $[]_-$ denotes the commutator of two operators, $S$ is the operator of total spin angular momentums. After performing a Fourier transformation, it yields
\begin{align}\label{SS_dynamics}
	\chi^{SS}(z=\omega+i\eta)=-\text{Im}\int^{\infty}_{-\infty}dte^{izt}\chi^{SS}(t).
\end{align}
We only display the imaginary parts since the real parts can be reproduced by the Kramers-Kronig relations. The dynamical susceptibilities $\chi^{LL}(\omega)$ and $\chi^{JJ}(\omega)$ for the corresponding orbital and total angular momentums can be obtained similarly. The calculation details regarding dynamical correlations are given in Appendix~\ref{ED-solver}. From Fig.~(\ref{susceptibility}) (a),
it is found that the low-energy excitations mainly come from spin fluctuations, while the high-energy parts are ascribed to the excitations of orbital momentum $\bf L$. Thus we focus on analyses of dynamical spin susceptibilities $\chi^{SS}(\omega)$ at $U=2.58$ for $\lambda=0.0$, $0.12$, and $0.14$. At $\lambda=0.0$, the system is in the OSM phase. The low-energy spin fluctuations consist of two parts, one is attributed to the Kondo screening of spins in $d_{yz/xz}$ bands by itinerant electrons, and the other is ascribed to the existence of spin triplets formed by $4$ electrons in $3$ $t_{2g}$ orbitals~\cite{SongPRB2017,KuglerPRB2019}.
At $\lambda=0.14$, the system becomes a Fermi liquid. The intensity of low-energy spin excitations is drastically enhanced. Since the low-energy spin excitations of $d_{yz/xz}$ bands remains almost unchanged for different $\lambda$ as seen in the inset of Fig.~\ref{susceptibility} (b), the abrupt increase should be ascribed to additional Kondo resonances between spins in $d_{xy}$ band and itinerant electrons. At $\lambda=0.12$, while electrons in $d_{xy}$ band become itinerant, the corresponding low-energy spin excitations are similar to those for $\lambda=0.0$, indicating that the spins in $d_{xy}$ band do not take part in the Kondo screening. The lack of Kondo screening in $d_{xy}$ band leads to the emergence of non-Fermi liquid. Hence we conclude that the OSnFL state is a result of orbitally-selective Kondo breakdown.

\section{discussion}
The calculated results show the Janus-faced influence of SOC on the Mott transitions in the presence of crystal field splitting at a filling of $n=2/3$. Owing to the orbital degeneracy lifted by crystal field splitting, the Mott transitions in different orbitals take place separately as $U$ increases. The opposite effect of SOC on the electronic correlations leads to the increase of $U_{xy}^c$ and decrease of $U_{yz/xz}^c$, which is in sharp contrast to previous theoretical results in the absence of crystal field splitting, where a common Mott transition occurs and the critical value of $U_c$ can only be increased or decreased by the SOC for given integer fillings~\cite{KimPRL2017,TrieblPRB2018,PiefkePRB2018}. Furthermore, the competition between SOC and Coulomb repulsion results in the OSnFL state, characterized by the coexistence of the Fermi liquid and non-Fermi liquid in different orbitals, which is fundamentally distinct from the case of $\lambda=0.0$, where the metallic ground state is always a Fermi liquid, as displayed in Fig.~\ref{phase_diagram}. The coexisting region is of particular importance as it provides a unique platform to study the breakdown of the Fermi-liquid picture and the nature of non-Fermi-liquid state. The appearance of OSnFL state requires remarkable orbital-differentiated correlations and sizable orbital hybridizations, reminiscent of non-Fermi liquid obtained within cluster DMFT~\cite{ZhangPRB2007,ParkPRL2008} where multiple sites are corresponding to multiple orbitals. In addition, we suppose that non-Fermi liquid might be the collective excitations of spin-charge separation.

Next, we discuss the relevance of our results to real materials. Ca$_{1.8}$Sr$_{0.2}$RuO$_4$ satisfies the conditions for the OSnFL state. On one hand, the intermediate SOC has been observed in Sr$_2$RuO$_4$~\cite{VeenstraPRL2014} and Ca$_2$RuO$_4$~\cite{FatuzzoPRB2015,GretarssonPRB2019}. On the other band, Ca$_{1.8}$Sr$_{0.2}$RuO$_4$ shows significant orbital-differentiated correlations with effective mass for $d_{xy}$ band much larger than that for $d_{yz/xz}$ bands~\cite{ShimoyamadaPRL2009,SutterPRB2019}. Furthermore, enhanced crystal field splitting induced by lattice distortion is detected in Ca$_{1.8}$Sr$_{0.2}$RuO$_4$~\cite{FriedtPRB2001}. Although the material-realistic DMFT calculations claimed that SOC hardly affects the correlations in Sr$_2$RuO$_4$~\cite{KimPRL2018} and Ca$_2$RuO$_4$~\cite{ZhangPRB2017}, we still suppose that Ca$_{1.8}$Sr$_{0.2}$RuO$_4$ should be susceptible to the SOC due to its proximity to the Mott transition. Hence orbitally-selective breakdown of the Fermi liquid observed experimentally in Ca$_{1.8}$Sr$_{0.2}$RuO$_4$~\cite{SutterPRB2019} may be a result of competition between SOC and orbital-differentiated correlations. Moreover, the non-Fermi liquid caused by strain in Ba$_2$RuO$_4$~\cite{BurganovPRL2016} and uniaxial pressure in Sr$_2$RuO$_4$~\cite{BarberPRL2018} may also be attributed to the competition, since both strain and uniaxial pressure slightly affect $d_{yz/xz}$ bands but significantly increase the renormalization mass of $d_{xy}$ band due to the band flattening, resulting in enhanced orbital differentiation. In addition, the OSnFL picture may have important implication for understanding the low-temperature non-Fermi liquid in CaRuO$_3$~\cite{LeePRB2002,KamalPRB2006,SchneiderPRL2014,YangPRB2016,LiuPRB2018}.

\section{conclusion}
In conclusion, we have investigated the three-orbital Hubbard model with both SOC and crystal field splitting using the DMFT combined with ED at $2/3$  filling. The OSM transitions take place as the orbital degeneracy is lifted by crystal field splitting in the absence of SOC. It is found that the SOC plays Janus-faced roles on the OSM transitions. While it suppresses the OSM transition at smaller $U$, it favors the OSM transition at larger $U$. The competition between the SOC and electronic correlations leads to the emergence of an OSnFL state, where the Fermi liquid coexits with non-Fermi liquid. The OSnFL state is originated from the orbitally-selective Kondo breakdown and can be applied to understand the exotic metals in $4d$ materials.

\begin{acknowledgments}
This work is financially supported by the National Natural Science Foundation of China (Grant No. 11774258, 12004283) and Postgraduate Education Reform Project of Tongji University (Grant No. GH1905), Z. Y. Song acknowledges the financial support by China Postdoctoral Science Foundation (Grant No. 2019M651563).
\end{acknowledgments}

\appendix
\section{Dynamical Mean Field Theory for Multi-orbital Hubbard Models}
\begin{figure*}
	\includegraphics[width=0.98\textwidth]{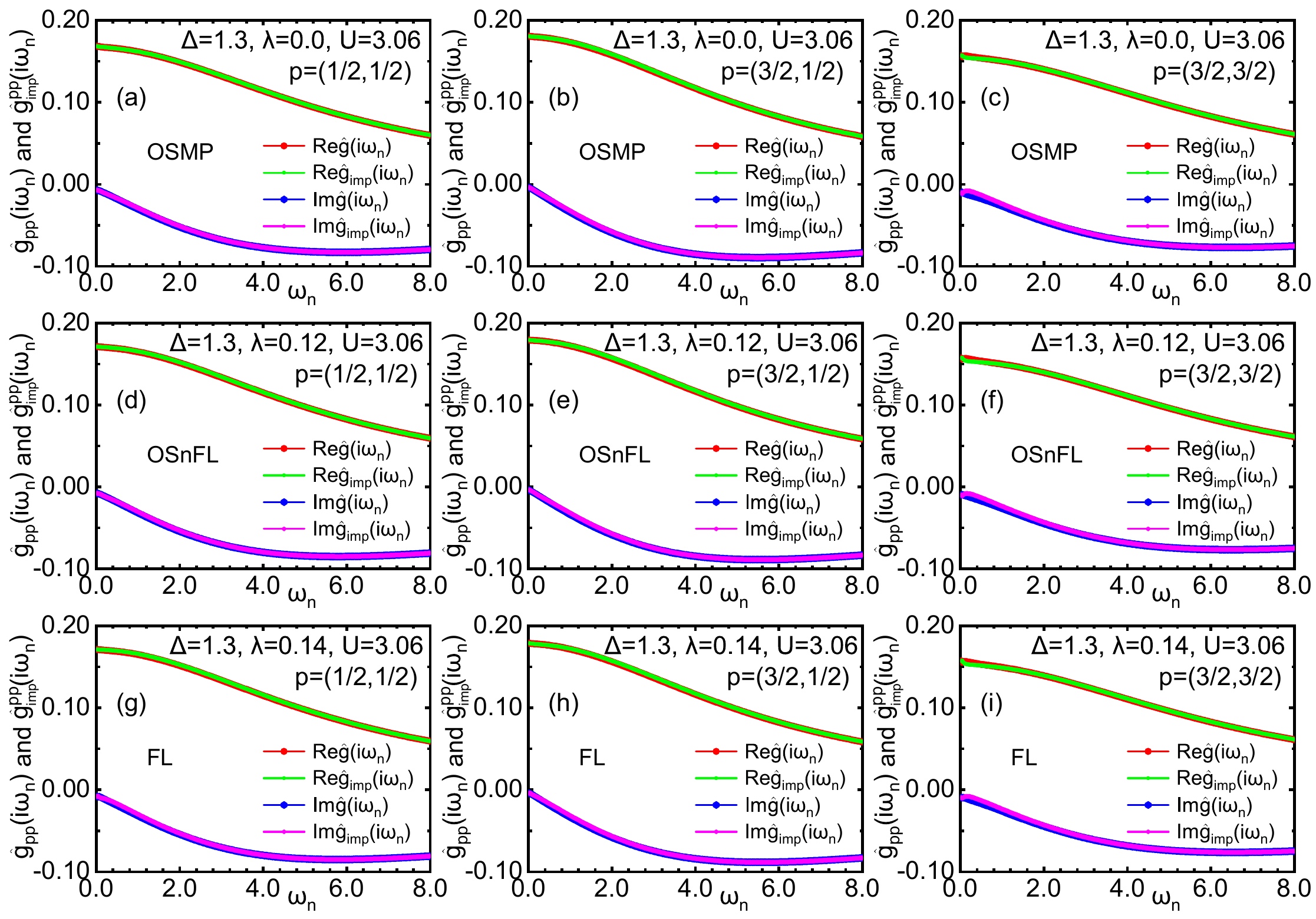}
	\caption{(color online) Comparison between the Weiss fields $\hat{g}(i\omega_n)$ and the noninteracting impurity Green's functions $\hat{g}_{\text{imp}}(i\omega_n)$ obtained from a discretized Anderson impurity model with 6 optimal bath sites. $\hat{g}(i\omega_n)$ can be well reproduced by $\hat{g}_{\text{imp}}(i\omega_n)$ in the OSM ((a)-(c)), the OSnFL ((d)-(f)), and the Fermi-liquid (FL) ((h)-(i)) phases for all $(J,m)$ orbitals including $p=(\frac{1}{2},\frac{1}{2})$, $p=(\frac{3}{2},\frac{1}{2})$, and $p=(\frac{3}{2},\frac{3}{2})$ orbitals. }
	\label{FitWeiss}
\end{figure*}

The DMFT~\cite{GeorgesRMP1996} was employed to investigated the multi-orbital Hubbard model~(\ref{Hubbardmodel}) on the bethe lattice with infinite coordination number, on which the DMFT has already been proved to be exact and the model  (\ref{Hubbardmodel})  can be exactly mapped onto an Anderson impurity model with self-consistent conditions $\hat{\Delta}(\omega)=\frac{D^2}{4}\hat{G}(\omega)$, where $\hat{G}(\omega)$  is the matrix of local lattice Green's functions, $\hat{\Delta}(\omega)$ denotes the matrix of hybridization functions  for the Anderson impurity model, and $D$ represents the half bandwidth. Note that the hat symbol is used to denote a matrix. When the spin-orbit coupling is considered, it is convenient to do the calculations in the $|J,\tau{m}\rangle$ basis,  where $J=\{\frac{1}{2}, \frac{3}{2}\}$, $m=\{\frac{1}{2},\frac{3}{2}\}$, $m\leq{J}$ and $\tau=\pm$ specifying a pair of Kramers doublet $m$, which are the eigenvectors of the Hamiltonian~(\ref{HSOC}).  In this representation, the Anderson impurity model reads
\begin{align}\label{Andersonmodel}
	H_{\text{imp}}&=\sum_{k\tau{m}}\epsilon_{km}C^{\dag}_{k\tau{m}}C_{k\tau{m}} +\sum_{p}\hat{E}_{p}d^{\dag}_{p}d_{p}+\sum_{pq}\hat{M}_{pq}d^{\dag}_pd_q\nonumber\\
	&+\sum_{kJ\tau{m}}V_{kJ\tau{m}}\left(C^{\dag}_{k\tau{m}}d_{J\tau{m}}+d^{\dag}_{J\tau{m}}C_{k\tau{m}}\right)\nonumber\\
	&+\frac{1}{4}\sum_{pqst}\tilde{U}^{pq}_{st}d^{\dag}_pd^{\dag}_qd_sd_t,
\end{align}	
where $\epsilon_{km}$ denotes the dispersion relationship of bath electrons, $p,q,s,t$ are the orbital indices of $(J,\tau{m})$, $\hat{E}_p$ is the eigenvalue of Hamiltonian~(\ref{HSOC}), $\hat{M}_{pq}$ is the matrix element of crystal field spliting in the $|J,\tau{m}\rangle$ basis, $V_{kJ\tau{m}}$ denotes the hybridization between local orbitals on the impurity and bath, and $\tilde{U}^{pq}_{st}=\sum_{\alpha\beta\gamma\delta}U_{\alpha\beta\gamma\delta}\hat{A}^*_{p\alpha}\hat{A}^*_{q\beta}\hat{A}_{s\gamma}\hat{A}_{t\delta}$ is the Coulomb interaction tensor element in the $|J,\tau{m}\rangle$ representation, where $U_{\alpha\beta\gamma\delta}$ is the Coulomb interaction tensor element in the $t_{2g}$ basis and $\hat{A}$ the unitary transformation~\cite{DuEPJB2013} between the $t_{2g}$ and $|J,\tau{m}\rangle$ basis.

The main tasks of present DMFT calculations are to solve the Anderson impurity model~(\ref{Andersonmodel}). We took ED as the impurity solver~\cite{GeorgesRMP1996}. In order to solve model~(\ref{Andersonmodel}) with ED, we have to use a finite number of bath sites to optimally fit the continuous bath. In this paper, totally six discrete bath sites were adopted to couple to three Kramers doublets, i.e. two bath sites per Kramers doublet. Owing to the presence of orbital mixing between $|\frac{1}{2},\frac{1}{2}\rangle$ ($|\frac{1}{2},-\frac{1}{2}\rangle$) and $|\frac{3}{2},\frac{1}{2}\rangle$ ($|\frac{3}{2},-\frac{1}{2}\rangle$) orbitals,  the corresponding cross-orbital hybridization functions, like $\hat{\Delta}_{\frac{1}{2},\frac{1}{2};\frac{3}{2},\frac{1}{2}}$ and $\hat{\Delta}_{\frac{1}{2},-\frac{1}{2};\frac{3}{2},-\frac{1}{2}}$, are considered.  From Fig.~\ref{FitWeiss}, it is found that the Weiss field $\hat{g}(i\omega_n)=(\hat{\Sigma}(i\omega_n)+\hat{G}^{-1}(i\omega_n))^{-1}$ with $\hat{\Sigma}(i\omega)$ the matrix of self-energy can be well reproduced by the noninteracting impurity Green's function $\hat{g}_{\text{imp}}(i\omega_n)$ defined by (\ref{Weiss_imp}) in the OSM ((a)-(c)), the OSnFL ((d)-(f)), and the Fermi-liquid ((g)-(i)) phases, suggesting that the finite size effects are negligible and the results presented in the paper are convincing. The procedures of the DMFT calculations  in combination with ED as the impurity solver are described below.

Starting from an initial set of bath parameters $\{\epsilon_{km},V_{kJ\tau{m}}\}$, we construct the Anderson model~(\ref{Andersonmodel}). The hybridization function of the corresponding impurity model~(\ref{Andersonmodel}) is written as
\begin{align}
\hat{\Delta}_{J,\tau{m};J^{\prime},\tau{m}}(i\omega_n)=	\sum_{k}\frac{V_{kJ\tau{m}}V_{kJ^{\prime}\tau{m}}}{i\omega_n-\epsilon_{km}},
\end{align}
where $\omega_n=\frac{(2n-1)\pi}{\beta}$ is the Matsubara frequency with a fictitious temperature $\beta{D}=200$, which serves as a low frequency cutoff. The noninteracting impurity Green's function of the Anderson model~(\ref{Andersonmodel}) is
\begin{align}\label{Weiss_imp}
\hat{g}_{\text{imp}}(i\omega_n)=i\omega_n+\mu-\hat{E}-\hat{M}-\hat{\Delta}^{-1}(i\omega_n).
\end{align}
After the impurity Green's function $\hat{G}_{\text{imp}}(i\omega_n)$ is obtained by solving model~(\ref{Andersonmodel}) with ED, the self-energy $\hat{\Sigma}(i\omega_n)$ can be calculated based on the Dyson's equation
\begin{align}
	\hat{\Sigma}(i\omega_n)=\hat{g}^{-1}_{\text{imp}}(i\omega_n)-\hat{G}^{-1}_{\text{imp}}(i\omega_n).
\end{align}	
Finally, the local Green's function of the three-orbital Hubbard model~(\ref{Hubbardmodel}) in the $|J,\tau{m}\rangle$ basis is calculated as
\begin{align}\label{LocalGreenFun}
	\hat{G}(i\omega_n)=\int^{+\infty}_{-\infty}\frac{\rho(\epsilon)d\epsilon}{i\omega_n+\mu-\hat{E}-\hat{M}-\hat{\Sigma}(i\omega_n)-\epsilon},
\end{align}
where $\rho(\omega)=\frac{2}{\pi{D}^2}\sqrt{D^2-\omega^2}$ is the density of states for three Kramers doublets, which is the same as the counterpart for the $t_{2g}$ orbitals.
Considering the self-consistent conditions $\hat{\Delta}(i\omega_n)=\frac{D^2}{4}\hat{G}(i\omega_n)$, we can iteratively calculate the local Green's function $\hat{G}(i\omega_n)$ through equations (\ref{Andersonmodel})---(\ref{LocalGreenFun}).

In our calculations, the parameters $\{\epsilon_{km},V_{kJ\tau{m}}\}$ to build the model (\ref{Andersonmodel}) are obtained by using the conjugate gradient method to minimizing the cost function
\begin{align}
	\chi=\frac{1}{N_{\text{max}}}\sum^{N_{\text{max}}}_{n=1}\frac{1}{\omega^2_n}\sum_{pq}\left|(\hat{g}(i\omega_n)-\hat{g}_{\text{imp}}(i\omega_n))_{pq}\right|,
\end{align}
where $N_{\text{max}}=256$ is the upper limit of the summation. Starting from a guessed Weiss field $\hat{g}(i\omega_n)$, we can self-consistently obtain the convergent results when the difference $\Delta_{g}$ between the new Weiss field $\hat{g}^{\text{new}}(i\omega_n)$ and the old Weiss field $\hat{g}^{\text{old}}(i\omega_n)$ is less than 10$^{-6}$. The difference $\Delta_{g}$ is defined as
\begin{align}
	\Delta_g=\text{max}\left\{\left|\hat{g}^{\text{new}}_{pq}(i\omega_n)-\hat{g}^{\text{old}}_{pq}(i\omega_n)\right|\right\}
\end{align}	

Although the present DMFT calculations were performed in the $|J,\tau{m}\rangle$ basis, the dynamical quantities, like local lattice Green's function $\hat{G}_{t_{2g}}(i\omega_n)$ and self-energy $\hat{\Sigma}_{t_{2g}}(\omega_n)$, in the $t_{2g}$ basis can be readly obtained via a unitary transformation
\begin{align}
	\hat{G}_{t_{2g}}(i\omega_n)=\hat{A}\hat{G}(i\omega_n)\hat{A}^{\dag},
\end{align}
and
\begin{align}
	\hat{\Sigma}_{t_{2g}}(i\omega_n)=\hat{A}\hat{\Sigma}(i\omega_n)\hat{A}^{\dag}.
\end{align}
Since ED has direct access to the real-frequency dynamical correlations as discribed in Appendix~\ref{ED-solver}, we can calculate the local lattice Green's function $\hat{G}(i\omega+i\eta)$ after the self-energy is obtained via the Dyson equation $\hat{\Sigma}(\omega+i\eta)=\hat{g}^{-1}_{\text{imp}}(\omega+i\eta)-\hat{G}^{-1}_{\text{imp}}(\omega+i\eta)$, where  the impurity Green's function $\hat{G}_{\text{imp}}(\omega+i\eta)$ is directly produced by ED. Similar to the Matsubara Green's function $\hat{G}_{t_{2g}}(i\omega_n)$, the local Green's function $\hat{G}_{t_{2g}}(\omega+i\eta)$ and self-energy $\hat{\Sigma}_{t_{2g}}(\omega+i\eta)$ in the $t_{2g}$ basis are calculated as
\begin{align}
	\hat{G}_{t_{2g}}(\omega+i\eta)=A\hat{G}(\omega+i\eta)A^{\dag},
\end{align}
and
\begin{align}
	\hat{\Sigma}_{t_{2g}}(\omega+i\eta)=A\hat{\Sigma}(\omega+i\eta)A^{\dag}.
\end{align}
On the basis of $\hat{G}_{t_{2g}}(\omega+i\eta)$, the projected density of states, as shown in the main text, is defined as
\begin{align}
	\rho_{\alpha\sigma}(\omega)=-\frac{1}{\pi}\text{Im}\left(\hat{G}_{t_{2g}}(\omega+i\eta)\right)_{\alpha\sigma,\alpha\sigma},
\end{align}
where $\alpha$ is the orbital index for $t_{2g}$ orbitals and $\sigma$ denotes electron spin.

\section{Exact Diagonalization}\label{ED-solver}
When ED is employed as impurity solver of the DMFT, it needs two steps to obtain the dynamical correlation functions. The first step is to calculate the ground-state energy $E_{g}$ and corresponding eigenvector $|\rangle$ of the Anderson impurity model~(\ref{Andersonmodel}) by the Lanczos method. On the basis of the Lanczos method, the $E_{g}$ and $|\rangle$ can be obtained by iteratively constructing a Krylov space $\{|\phi_n\rangle\}$ from an arbitrary initial configuration $|\phi_1\rangle$ via
\begin{align}
	|\tilde\phi_{n+1}\rangle=H_{\text{imp}}|\phi_n\rangle-a_n|\phi_n\rangle-b^2_{n}|\phi_{n-1}\rangle\label{Lanczos_iter},
\end{align}
where n= 2, 3, 4, $\cdots$, $a_n=\langle\phi_n|H_{\text{imp}}|\phi_n\rangle$, $b^2_n=\langle\tilde\phi_{n+1}|\tilde\phi_{n+1}\rangle$. Note $|\phi_n\rangle$ denotes a normalized vector, $b_1=0$, $|\phi_0\rangle=0$, $|\tilde\phi_2\rangle=H_{\text{imp}}|\phi_1\rangle-a_1|\phi_1\rangle$ and $a_1=\langle\phi_1|H_{\text{imp}}|\phi_1\rangle$. The iteration~(\ref*{Lanczos_iter}) continues until $b_n$ is less than a threshold. In this basis, the Hamiltonian for the Anderson impurity model~(\ref{Andersonmodel}) is a tridiagonal matrix and simply reads
\begin{align}
	\begin{bmatrix}
		a_1 & b_2 & 0      &0     & \cdots\\
		b_2& a_2 &  b_3 &0     & \cdots\\
		0    & b_3 & a_3 & b_4 &\cdots\\
		0   & 0     & b_4  & a_4 &\cdots \\
		\vdots&\vdots&\vdots&\vdots
	\end{bmatrix},
\end{align}
which can be diagonalized by the modern standard  library subroutines.

The second step is to calculate the dynamical correlation function based on the ground-state energy $E_g$ and ground-state eigenvector $|\rangle$ obtained in the first step. For the given operator $O$, the real-time dynamical correlation function is defined as
\begin{align}
	C_{\alpha\beta}(t)=-i\theta(t)\langle|\left[O_{\alpha}(t),O_{\beta}\right]_{\xi}|\rangle	
\end{align}
where $\theta(t)$ is a step function of time $t$, $O_{\alpha}(t)=e^{iH_{\text{imp}}t}O_{\alpha}e^{-iH_{\text{imp}}t}$, $[]_{\xi}$ denotes the commutator of two operators, $\xi=1$ if $O$ is a Fermi operator and $\xi=-1$ otherwise. After performing a Fourier transformation, the dynamical correlation function on the real frequency axis can be written as
\begin{align}
	C_{\alpha\beta}(\omega+i\eta)=C^{>}_{\alpha\beta}(\omega+i\eta)	+\xi{}C^{<}_{\alpha\beta}(\omega+i\eta)\label{DCF_real},
\end{align}
where $\eta$ is a broadening factor and
\begin{align}
	C^{>}_{\alpha\beta}(\omega+i\eta)&=\langle|O_{\alpha}\frac{1}{\omega-H_{\text{imp}}+E_g+i\eta}O_{\beta}|\rangle,\label{DCF_elec}\\
	C^{<}_{\alpha\beta}(\omega+i\eta)&=\langle|O_{\beta}\frac{1}{\omega+H_{\text{imp}}-E_g+i\eta}O_{\alpha}|\rangle.	
\end{align}	
Similar to calculate the ground-state energy $E_g$ and eigenvector $|\rangle$, the method of the Krylov space can be applied to calculate both dynamical correlation functions above.

In order to calculate $C^{>}_{\alpha\beta}(\omega+i\eta)$, we start the Lanczos iteratons with the initial vector $|\phi_1^{\beta}\rangle=O_{\beta}|\rangle/\langle|O^{\dag}_{\beta}O_{\beta}|\rangle$ to construct the new basis $\{|\phi^{\beta}_n\rangle\}$. By inserting the completeness $\sum_n|\phi^{\beta}_n\rangle\langle\phi^{\beta}_n|=1$ into equation~(\ref{DCF_elec}), it yields
\begin{align}
	C^{>}_{\alpha\beta}(\omega+i\eta)&=\sqrt{\langle|O_{\alpha}O^{\dag}_{\alpha}|\rangle}\sqrt{\langle|O^{\dag}_{\beta}O_{\beta}|\rangle}\nonumber\\
	&\times\sum_nU^{\alpha\beta}_nV^{\beta}_n(\omega+i\eta)\label{DCF_elec2}
\end{align}
where
\begin{align}
V^{\beta}_{n}(\omega+i\eta)=\langle\phi^{\beta}_n|\frac{1}{\omega-H_{\text{imp}}+E_g+i\eta}|\phi^{\beta}_1\rangle,
\end{align}
and $U^{\alpha\beta}_n=\langle\phi^{\alpha}_1|\phi^{\beta}_n\rangle$ with $\langle\phi^{\alpha}_1|=\langle|O_{\alpha}/\sqrt{\langle|O_{\alpha}O^{\dag}_{\alpha}|\rangle}$. It is obvious that the main difficulties are to calculate $V^{\beta}_{n}(\omega+i\eta)$ in the new basis. By taking advantage of the completeness of $\sum_n|\phi^{\beta}_n\rangle\langle\phi^{\beta}_n|=1$  and the identity
\begin{align}
	\langle\phi^{\beta}_m|(\omega-H_{\text{imp}}+E_g)\frac{1}{\omega-H_{\text{imp}}+E_g}	|\phi^{\beta}_1\rangle=\delta_{m,1},
\end{align}
$V^{\beta}_{n}(\omega+i\eta)$ can be obtained through
\begin{align}
	S^{\beta}_{mn}(\omega+i\eta)V^{\beta}_n(\omega+i\eta)=E_m,	\label{SV_LinearEquation}
\end{align}
where $E_m=\delta_{m,1}$ and $S^{\beta}_{mn}(\omega+i\eta)$ is a tridiagonal matrix
\begin{widetext}
	\begin{align}
		S^{\beta}_{mn}(\omega+i\eta)=
		\begin{bmatrix}
			\omega-a_1+E_g+i\eta & -b_2 & 0      &0     & \cdots\\
			-b_2& \omega-a_2+E_g+i\eta &  -b_3 &0     & \cdots\\
			0    & -b_3 & \omega-a_3+E_g+i\eta & -b_4 &\cdots\\
			0   & 0     & -b_4  & \omega-a_4+E_g+i\eta &\cdots \\
			\vdots&\vdots&\vdots&\vdots
		\end{bmatrix},
	\end{align}
\end{widetext}
where $a_n$ and $b_n$ are the coefficients generated by the Lanczos iterations with the initial configuration $|\phi^{\beta}_1\rangle$.  The linear equations~(\ref{SV_LinearEquation}) can be easily solved by the standard library subroutines. Finally, we can obtain the dynamical correlation function $C^{>}_{\alpha\beta}(\omega+i\eta)$ by solving equations~(\ref{DCF_elec2}) and~(\ref{SV_LinearEquation}). As for $C^{<}_{\alpha\beta}(\omega+i\eta)$, it can be obtained similarly.

When substituting the Matsubara frequency $i\omega_n$ for the real frequency $\omega+i\eta$ in equation~(\ref{DCF_real}), the dynamical correlation functions $C_{\alpha\beta}(i\omega_n)$ on the Matsubara frequency axis can be obtained by following the above two-step procedures. It suggests ED has direct access to the dynamical quantities on both real and Matsubara frequency axes. This is in sharp contrast to the quantum Monte Carlo, which cannot directly sample the real-frequency dynamical correlations.

\section{Results with Eight Bath Sites Coupled to Three Kramers Doublets}

\begin{figure}[htbp]
	\includegraphics[width=0.48\textwidth]{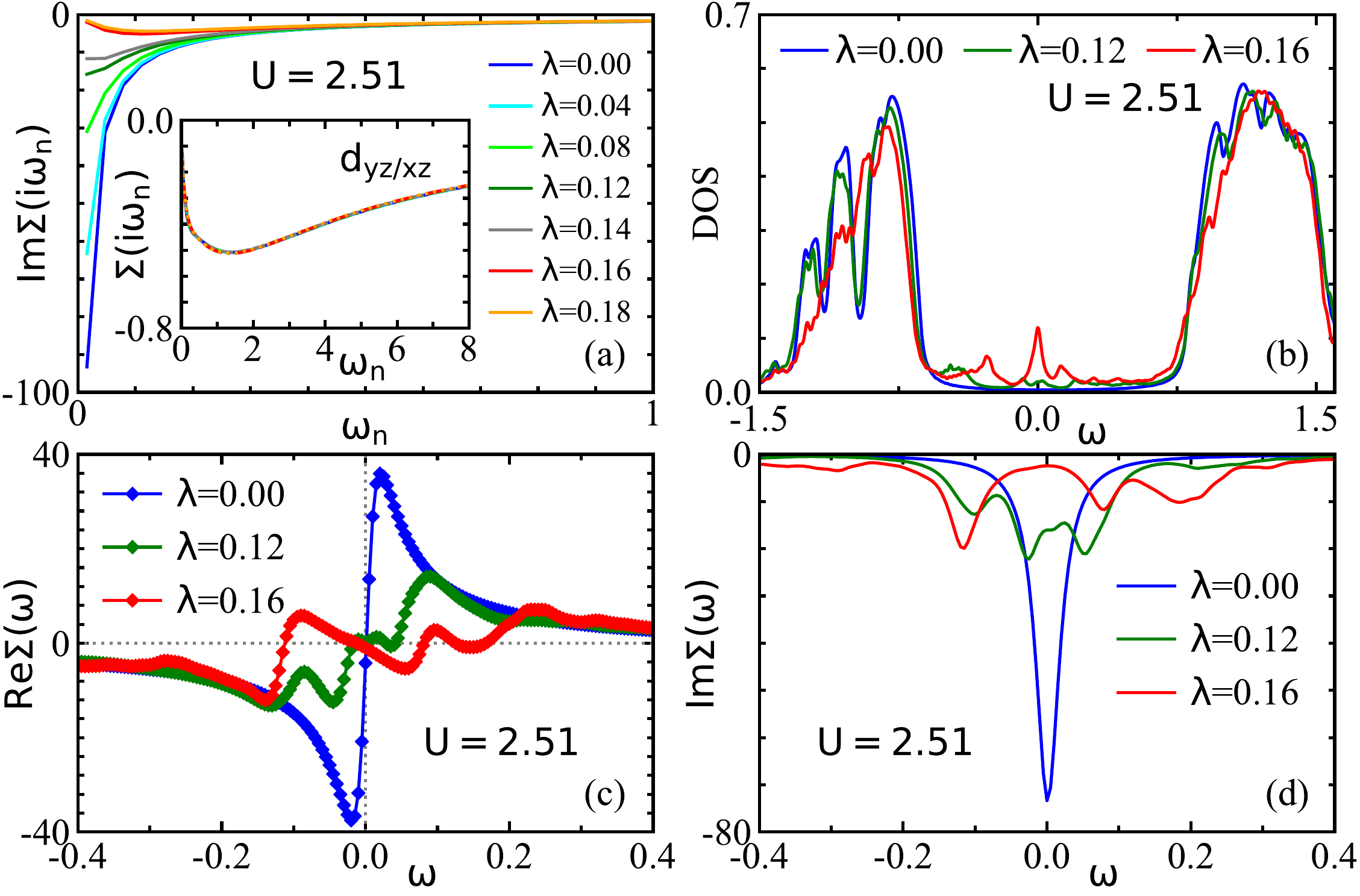}
	\caption{(color online) The influence of spin-orbit coupling on the imaginary part of Matsubara self-energy for $d_{xy}$ band, those for $d_{yz/xz}$ bands are shown in the inset. (b) Density of states for $d_{xy}$ band in the Mott ($\lambda=0.0$), non-Fermi-liquid ($\lambda=0.12$) and the Fermi-liquid ($\lambda=0.16$) states. (c) Real and (d) imaginary part of self-energy of $d_{xy}$ band on the real frequency axis. Here, the electron repulsion is fixed at $U=2.51$.}
	\label{OSnFL_8bath}
\end{figure}

In order to further investigate the effect of bath discretization on the OSnFL phase, we have done the DMFT calculations with eight bath sites at zero temperature using ED as the impurity solver, where four bath sites couple to $|\frac{1}{2},\pm\frac{1}{2}\rangle$ and $|\frac{3}{2},\pm\frac{1}{2}\rangle$ Kramers doublets and the remained four bath sites couple to $|\frac{3}{2},\pm\frac{3}{2}\rangle$ doublet. The calculated results at $U=2.51$ are displayed in Fig.~\ref{OSnFL_8bath}, which are similar to those obtained with six bath sites. The OSnFL state can be clearly identified by the imaginary part of self-energy $\textrm{Im}\Sigma(i\omega_n)$ on the Matsubara frequency axis. As shown in the inset of Fig.~\ref{OSnFL_8bath} (a), $\textrm{Im}\Sigma_{yz/xz}(i\omega_n)$ for $d_{yz/xz}$ orbitals always extrapolates to zero when $\omega_n$ goes to zero, suggesting the typical Fermi-liquid behavior in $d_{yz/xz}$ orbitals. In contrast, at around $\lambda\approx0.12$, $\textrm{Im}\Sigma_{xy}(i\omega_n)$ for $d_{xy}$ orbital approaches a finite value at low frequencies, which is distinctly different from the divergent behavior for $\lambda<0.8$ and the tendency to approach zero for $\lambda>0.14$, indicating the occurrence of the Fermi-liquid breakdown in $d_{xy}$ orbital due to the finite scattering rate at zero temperature. 

The self-energy $\Sigma_{xy}(\omega)$ on the real frequency axis further manifest the SOC-induced non-Fermi-liquid nature of $d_{xy}$ orbital. Fig.~\ref{OSnFL_8bath} (c) and (d) shows the real part of self-energy Re$\Sigma_{xy}(\omega)$ and the imaginary part of self-energy Im$\Sigma_{xy}(\omega)$ at $U=2.51$ for $\lambda=0.0$, $\lambda=0.12$ and $\lambda=0.16$.  For the case of $\lambda=0.12$, the positive slop of Re$\Sigma_{xy}(\omega)$ and finite value of Im$\Sigma_{xy}(\omega)$ at the Fermi level indicate the breakdown of the Fermi liquid at zero temperature. In contrast, the sharp slop of Re$\Sigma_{xy}(\omega)$ and divergent Im$\Sigma_{xy}(\omega)$ at $\omega=0$ suggest a Mott-insulating state for $\lambda=0.0$ case, and the linearly $\omega-$dependent Re$\Sigma_{xy}(\omega)$ and quadratically $\omega^2-$ dependent Im$\Sigma_{xy}(\omega)$  in the vicinity of the Fermi level indicate a typical Fermi-liquid behavior for $\lambda=0.16$ case. Fig.~\ref{OSnFL_8bath} (c) displays the density of states for $d_{xy}$ orbital in the Mott, non-Fermi-liquid and Fermi-liquid states at $\lambda=0.0$, $0.12$ and $0.16$, respectively.

\begin{figure}[b]
	\includegraphics[width=0.48\textwidth]{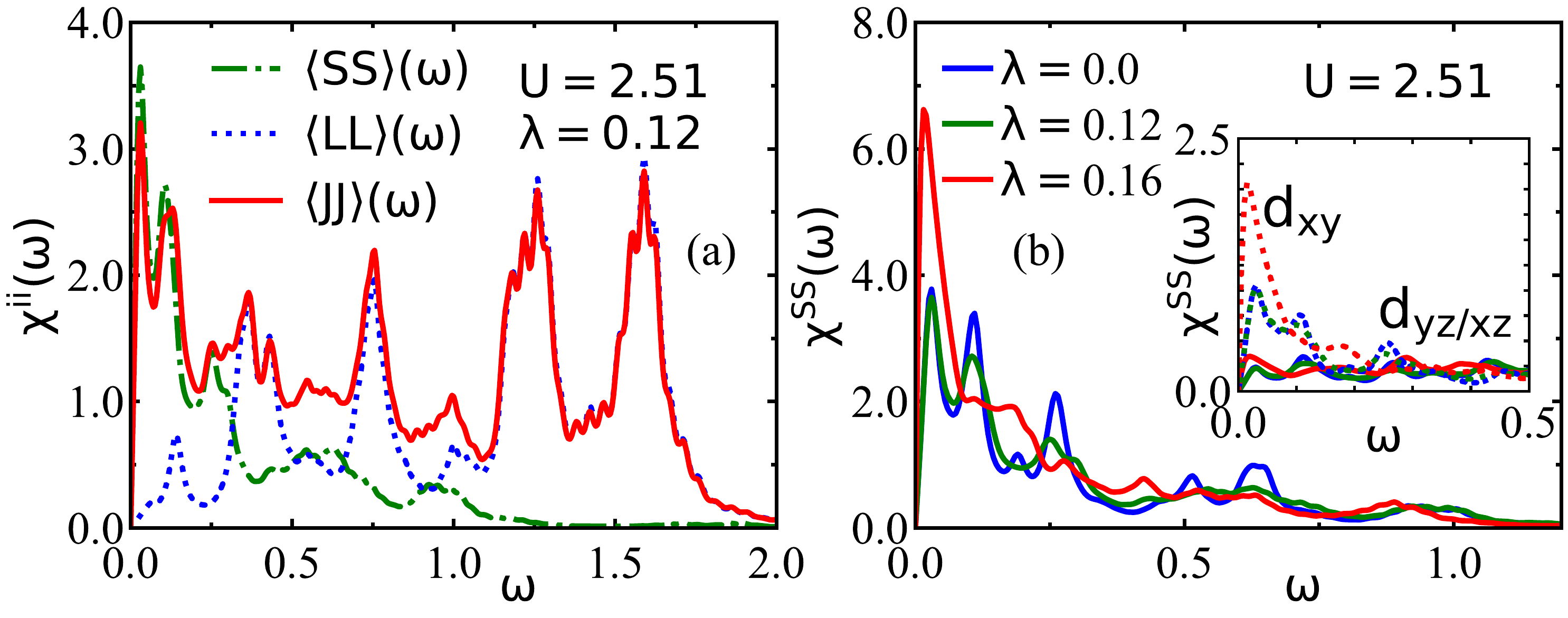}
	\caption{(color online) (a) Dynamical susceptibilities for spin ($\bf S$), orbital ($\bf L$) and total ($\bf J$) angular momentums at $U=2.51$ and $\lambda=0.12$. (b) Dynamical spin susceptibilities at $U=2.51$ for $\lambda=0.0$, $\lambda=0.12$ and $\lambda=0.16$ cases, the inset shows corresponding orbitally-resolved dynamical spin susceptibilities, where solid lines denote $d_{yz/xz}$ orbitals and dashed lines $d_{xy}$ orbital. }
	\label{dynamics_8bath}
\end{figure}

In order to reveal the mechanism which is responsible for the appearance of the OSnFL state, we have calculated the dynamical susceptibilities $\chi^{ii}(\omega)$, defined by Eq (\ref*{SS_correlation}) and (\ref*{SS_dynamics}), for total spin momentum $\mathbf{S}$, total orbital angular momentum $\mathbf{L}$ and total angular momentum $\mathbf{J}$ at $U=2.51$ and $\lambda=0.12$, as displayed in Fig.~\ref{dynamics_8bath} (a). This result is similar to that obtained with totally six bath sites. It is obvious that the low-energy excitations of the total angular momentum $\mathbf{J}$ are mainly contributed by the spin fluctuations and the high-energy ones are ascribed to the excitations of the total orbital angular momentum $\mathbf{L}$. Since the low-energy excitations were found to dominate the physical properties of the correlated multiorbital systems~\cite{SongPRB2017,KuglerPRB2019}, we now pay attention on the low-energy spin fluctuations. Fig.~\ref{dynamics_8bath} (b) depicts the spin susceptibilities at $U=2.51$ for $\lambda=0.0$, $0.12$ and $0.16$ cases. As explained in the maintext, at $\lambda=0.0$, the system is in the OSM state. The appearance of low-energy spin excitations is due to the spins in $d_{yz/xz}$ orbitals screened by their itinerant electrons and the formation of the local triplets with three $t_{2g}$ orbitals filled by four electrons. At $\lambda=0.16$, the abrupt jump in spin excitations is attributed to the presence of the additional Kondo resonances in $d_{xy}$ orbital because the spin susceptibilities for $d_{yz/xz}$ orbitals keep almost unchange, as displayed in the inset of Fig.~\ref{dynamics_8bath} (b). In contrast, at $\lambda=0.12$, the spin susceptibilities are almost the same with those at $\lambda=0.0$, indicating the spins in $d_{xy}$ band do not participate in the Kondo screening. Obviously, the lack of the Kondo resonances in $d_{xy}$ orbital leads to the non-Fermi liquid behavior.

In summary, the results calculated with eight bath sites are qualitatively consistent with those obtained with six bath sites, suggesting the finite-size effect is negligible.

\bibliography{OSMPSOC_bb}

\providecommand{\noopsort}[1]{}\providecommand{\singleletter}[1]{#1}%
\begin{thebibliography}{67}%
\makeatletter
\providecommand \@ifxundefined [1]{%
 \@ifx{#1\undefined}
}%
\providecommand \@ifnum [1]{%
 \ifnum #1\expandafter \@firstoftwo
 \else \expandafter \@secondoftwo
 \fi
}%
\providecommand \@ifx [1]{%
 \ifx #1\expandafter \@firstoftwo
 \else \expandafter \@secondoftwo
 \fi
}%
\providecommand \natexlab [1]{#1}%
\providecommand \enquote  [1]{``#1''}%
\providecommand \bibnamefont  [1]{#1}%
\providecommand \bibfnamefont [1]{#1}%
\providecommand \citenamefont [1]{#1}%
\providecommand \href@noop [0]{\@secondoftwo}%
\providecommand \href [0]{\begingroup \@sanitize@url \@href}%
\providecommand \@href[1]{\@@startlink{#1}\@@href}%
\providecommand \@@href[1]{\endgroup#1\@@endlink}%
\providecommand \@sanitize@url [0]{\catcode `\\12\catcode `\$12\catcode
  `\&12\catcode `\#12\catcode `\^12\catcode `\_12\catcode `\%12\relax}%
\providecommand \@@startlink[1]{}%
\providecommand \@@endlink[0]{}%
\providecommand \url  [0]{\begingroup\@sanitize@url \@url }%
\providecommand \@url [1]{\endgroup\@href {#1}{\urlprefix }}%
\providecommand \urlprefix  [0]{URL }%
\providecommand \Eprint [0]{\href }%
\providecommand \doibase [0]{http://dx.doi.org/}%
\providecommand \selectlanguage [0]{\@gobble}%
\providecommand \bibinfo  [0]{\@secondoftwo}%
\providecommand \bibfield  [0]{\@secondoftwo}%
\providecommand \translation [1]{[#1]}%
\providecommand \BibitemOpen [0]{}%
\providecommand \bibitemStop [0]{}%
\providecommand \bibitemNoStop [0]{.\EOS\space}%
\providecommand \EOS [0]{\spacefactor3000\relax}%
\providecommand \BibitemShut  [1]{\csname bibitem#1\endcsname}%
\let\auto@bib@innerbib\@empty
\bibitem [{\citenamefont {Kamihara}\ \emph {et~al.}(2006)\citenamefont
  {Kamihara}, \citenamefont {Hiramatsu}, \citenamefont {Hirano}, \citenamefont
  {Kawamura}, \citenamefont {Yanagi}, \citenamefont {Kamiya},\ and\
  \citenamefont {Hosono}}]{KamiharaJACS2006}%
  \BibitemOpen
  \bibfield  {author} {\bibinfo {author} {\bibfnamefont {Yoichi}\ \bibnamefont
  {Kamihara}}, \bibinfo {author} {\bibfnamefont {Hidenori}\ \bibnamefont
  {Hiramatsu}}, \bibinfo {author} {\bibfnamefont {Masahiro}\ \bibnamefont
  {Hirano}}, \bibinfo {author} {\bibfnamefont {Ryuto}\ \bibnamefont
  {Kawamura}}, \bibinfo {author} {\bibfnamefont {Hiroshi}\ \bibnamefont
  {Yanagi}}, \bibinfo {author} {\bibfnamefont {Toshio}\ \bibnamefont {Kamiya}},
  \ and\ \bibinfo {author} {\bibfnamefont {Hideo}\ \bibnamefont {Hosono}},\
  }\bibfield  {title} {\enquote {\bibinfo {title} {{Iron-Based Layered
  Superconductor:{\thinspace} LaOFeP}},}\ }\href {\doibase 10.1021/ja063355c}
  {\bibfield  {journal} {\bibinfo  {journal} {J. Am. Chem. Soc.}\ }\textbf
  {\bibinfo {volume} {128}},\ \bibinfo {pages} {10012} (\bibinfo {year}
  {2006})}\BibitemShut {NoStop}%
\bibitem [{\citenamefont {Kamihara}\ \emph {et~al.}(2008)\citenamefont
  {Kamihara}, \citenamefont {Watanabe}, \citenamefont {Hirano},\ and\
  \citenamefont {Hosono}}]{KamiharaJACS2008}%
  \BibitemOpen
  \bibfield  {author} {\bibinfo {author} {\bibfnamefont {Yoichi}\ \bibnamefont
  {Kamihara}}, \bibinfo {author} {\bibfnamefont {Takumi}\ \bibnamefont
  {Watanabe}}, \bibinfo {author} {\bibfnamefont {Masahiro}\ \bibnamefont
  {Hirano}}, \ and\ \bibinfo {author} {\bibfnamefont {Hideo}\ \bibnamefont
  {Hosono}},\ }\bibfield  {title} {\enquote {\bibinfo {title} {{Iron-Based
  Layered Superconductor La[O$_{1-x}$F$_x$]FeAs (x=0.05-0.12) with Tc = 26
  K}},}\ }\href {\doibase 10.1021/ja800073m} {\bibfield  {journal} {\bibinfo
  {journal} {J. Am. Chem. Soc.}\ }\textbf {\bibinfo {volume} {130}},\ \bibinfo
  {pages} {3296} (\bibinfo {year} {2008})}\BibitemShut {NoStop}%
\bibitem [{\citenamefont {Witczak-Krempa}\ \emph {et~al.}(2014)\citenamefont
  {Witczak-Krempa}, \citenamefont {Chen}, \citenamefont {Kim},\ and\
  \citenamefont {Balents}}]{WitczakARCMP2014}%
  \BibitemOpen
  \bibfield  {author} {\bibinfo {author} {\bibfnamefont {William}\ \bibnamefont
  {Witczak-Krempa}}, \bibinfo {author} {\bibfnamefont {Gang}\ \bibnamefont
  {Chen}}, \bibinfo {author} {\bibfnamefont {Yong~Baek}\ \bibnamefont {Kim}}, \
  and\ \bibinfo {author} {\bibfnamefont {Leon}\ \bibnamefont {Balents}},\
  }\bibfield  {title} {\enquote {\bibinfo {title} {{Correlated Quantum
  Phenomena in the Strong Spin-Orbit Regime}},}\ }\href {\doibase
  10.1146/annurev-conmatphys-020911-125138} {\bibfield  {journal} {\bibinfo
  {journal} {Annu. Rev. Condens. Matter Phys.}\ }\textbf {\bibinfo {volume}
  {5}},\ \bibinfo {pages} {57} (\bibinfo {year} {2014})}\BibitemShut {NoStop}%
\bibitem [{\citenamefont {Rau}\ \emph {et~al.}(2016)\citenamefont {Rau},
  \citenamefont {Lee},\ and\ \citenamefont {Kee}}]{RauARCMP2016}%
  \BibitemOpen
  \bibfield  {author} {\bibinfo {author} {\bibfnamefont {Jeffrey~G.}\
  \bibnamefont {Rau}}, \bibinfo {author} {\bibfnamefont {Eric Kin-Ho}\
  \bibnamefont {Lee}}, \ and\ \bibinfo {author} {\bibfnamefont {Hae-Young}\
  \bibnamefont {Kee}},\ }\bibfield  {title} {\enquote {\bibinfo {title}
  {{Spin-Orbit Physics Giving Rise to Novel Phases in Correlated Systems:
  Iridates and Related Materials}},}\ }\href {\doibase
  10.1146/annurev-conmatphys-031115-011319} {\bibfield  {journal} {\bibinfo
  {journal} {Annu. Rev. Condens. Matter Phys.}\ }\textbf {\bibinfo {volume}
  {7}},\ \bibinfo {pages} {195} (\bibinfo {year} {2016})}\BibitemShut {NoStop}%
\bibitem [{\citenamefont {Martins}\ \emph {et~al.}(2017)\citenamefont
  {Martins}, \citenamefont {Aichhorn},\ and\ \citenamefont
  {Biermann}}]{MartinsJPCM2017}%
  \BibitemOpen
  \bibfield  {author} {\bibinfo {author} {\bibfnamefont {C.}~\bibnamefont
  {Martins}}, \bibinfo {author} {\bibfnamefont {M.}~\bibnamefont {Aichhorn}}, \
  and\ \bibinfo {author} {\bibfnamefont {S.}~\bibnamefont {Biermann}},\
  }\bibfield  {title} {\enquote {\bibinfo {title} {{Coulomb correlations in 4d
  and 5d oxides from first principles{\textemdash}or how spin{\textendash}orbit
  materials choose their effective orbital degeneracies}},}\ }\href {\doibase
  10.1088/1361-648x/aa648f} {\bibfield  {journal} {\bibinfo  {journal} {Journal
  of Physics: Condensed Matter}\ }\textbf {\bibinfo {volume} {29}},\ \bibinfo
  {pages} {263001} (\bibinfo {year} {2017})}\BibitemShut {NoStop}%
\bibitem [{\citenamefont {Meng}\ \emph {et~al.}(2014)\citenamefont {Meng},
  \citenamefont {Kim},\ and\ \citenamefont {Kee}}]{MengPRL2014}%
  \BibitemOpen
  \bibfield  {author} {\bibinfo {author} {\bibfnamefont {Zi~Yang}\ \bibnamefont
  {Meng}}, \bibinfo {author} {\bibfnamefont {Yong~Baek}\ \bibnamefont {Kim}}, \
  and\ \bibinfo {author} {\bibfnamefont {Hae-Young}\ \bibnamefont {Kee}},\
  }\bibfield  {title} {\enquote {\bibinfo {title} {{Odd-Parity Triplet
  Superconducting Phase in Multiorbital Materials with a Strong Spin-Orbit
  Coupling: Application to Doped ${\mathrm{Sr}}_{2}{\mathrm{IrO}}_{4}$}},}\
  }\href {\doibase 10.1103/PhysRevLett.113.177003} {\bibfield  {journal}
  {\bibinfo  {journal} {Phys. Rev. Lett.}\ }\textbf {\bibinfo {volume} {113}},\
  \bibinfo {pages} {177003} (\bibinfo {year} {2014})}\BibitemShut {NoStop}%
\bibitem [{\citenamefont {Chaloupka}\ and\ \citenamefont
  {Khaliullin}(2016)}]{ChaloupkaPRL2016}%
  \BibitemOpen
  \bibfield  {author} {\bibinfo {author} {\bibfnamefont {Ji\ifmmode
  \check{r}\else~\v{r}\fi{}\'{\i}}\ \bibnamefont {Chaloupka}}\ and\ \bibinfo
  {author} {\bibfnamefont {Giniyat}\ \bibnamefont {Khaliullin}},\ }\bibfield
  {title} {\enquote {\bibinfo {title} {{Doping-Induced Ferromagnetism and
  Possible Triplet Pairing in ${d}^{4}$ Mott Insulators}},}\ }\href {\doibase
  10.1103/PhysRevLett.116.017203} {\bibfield  {journal} {\bibinfo  {journal}
  {Phys. Rev. Lett.}\ }\textbf {\bibinfo {volume} {116}},\ \bibinfo {pages}
  {017203} (\bibinfo {year} {2016})}\BibitemShut {NoStop}%
\bibitem [{\citenamefont {Kim}\ \emph {et~al.}(2009)\citenamefont {Kim},
  \citenamefont {Ohsumi}, \citenamefont {Komesu}, \citenamefont {Sakai},
  \citenamefont {Morita}, \citenamefont {Takagi},\ and\ \citenamefont
  {Arima}}]{KimScience2009}%
  \BibitemOpen
  \bibfield  {author} {\bibinfo {author} {\bibfnamefont {B.~J.}\ \bibnamefont
  {Kim}}, \bibinfo {author} {\bibfnamefont {H.}~\bibnamefont {Ohsumi}},
  \bibinfo {author} {\bibfnamefont {T.}~\bibnamefont {Komesu}}, \bibinfo
  {author} {\bibfnamefont {S.}~\bibnamefont {Sakai}}, \bibinfo {author}
  {\bibfnamefont {T.}~\bibnamefont {Morita}}, \bibinfo {author} {\bibfnamefont
  {H.}~\bibnamefont {Takagi}}, \ and\ \bibinfo {author} {\bibfnamefont
  {T.}~\bibnamefont {Arima}},\ }\bibfield  {title} {\enquote {\bibinfo {title}
  {{Phase-Sensitive Observation of a Spin-Orbital Mott State in
  Sr$_2$IrO$_4$}},}\ }\href {\doibase 10.1126/science.1167106} {\bibfield
  {journal} {\bibinfo  {journal} {Science}\ }\textbf {\bibinfo {volume}
  {323}},\ \bibinfo {pages} {1329} (\bibinfo {year} {2009})}\BibitemShut
  {NoStop}%
\bibitem [{\citenamefont {Chaloupka}\ \emph {et~al.}(2010)\citenamefont
  {Chaloupka}, \citenamefont {Jackeli},\ and\ \citenamefont
  {Khaliullin}}]{ChaloupkaPRL2010}%
  \BibitemOpen
  \bibfield  {author} {\bibinfo {author} {\bibfnamefont {Ji\ifmmode
  \check{r}\else~\v{r}\fi{}\'{\i}}\ \bibnamefont {Chaloupka}}, \bibinfo
  {author} {\bibfnamefont {George}\ \bibnamefont {Jackeli}}, \ and\ \bibinfo
  {author} {\bibfnamefont {Giniyat}\ \bibnamefont {Khaliullin}},\ }\bibfield
  {title} {\enquote {\bibinfo {title} {{Kitaev-Heisenberg Model on a Honeycomb
  Lattice: Possible Exotic Phases in Iridium Oxides
  ${A}_{2}{\mathrm{IrO}}_{3}$}},}\ }\href {\doibase
  10.1103/PhysRevLett.105.027204} {\bibfield  {journal} {\bibinfo  {journal}
  {Phys. Rev. Lett.}\ }\textbf {\bibinfo {volume} {105}},\ \bibinfo {pages}
  {027204} (\bibinfo {year} {2010})}\BibitemShut {NoStop}%
\bibitem [{\citenamefont {Comin}\ \emph {et~al.}(2012)\citenamefont {Comin},
  \citenamefont {Levy}, \citenamefont {Ludbrook}, \citenamefont {Zhu},
  \citenamefont {Veenstra}, \citenamefont {Rosen}, \citenamefont {Singh},
  \citenamefont {Gegenwart}, \citenamefont {Stricker}, \citenamefont {Hancock},
  \citenamefont {van~der Marel}, \citenamefont {Elfimov},\ and\ \citenamefont
  {Damascelli}}]{CominPRL2012}%
  \BibitemOpen
  \bibfield  {author} {\bibinfo {author} {\bibfnamefont {R.}~\bibnamefont
  {Comin}}, \bibinfo {author} {\bibfnamefont {G.}~\bibnamefont {Levy}},
  \bibinfo {author} {\bibfnamefont {B.}~\bibnamefont {Ludbrook}}, \bibinfo
  {author} {\bibfnamefont {Z.-H.}\ \bibnamefont {Zhu}}, \bibinfo {author}
  {\bibfnamefont {C.~N.}\ \bibnamefont {Veenstra}}, \bibinfo {author}
  {\bibfnamefont {J.~A.}\ \bibnamefont {Rosen}}, \bibinfo {author}
  {\bibfnamefont {Yogesh}\ \bibnamefont {Singh}}, \bibinfo {author}
  {\bibfnamefont {P.}~\bibnamefont {Gegenwart}}, \bibinfo {author}
  {\bibfnamefont {D.}~\bibnamefont {Stricker}}, \bibinfo {author}
  {\bibfnamefont {J.~N.}\ \bibnamefont {Hancock}}, \bibinfo {author}
  {\bibfnamefont {D.}~\bibnamefont {van~der Marel}}, \bibinfo {author}
  {\bibfnamefont {I.~S.}\ \bibnamefont {Elfimov}}, \ and\ \bibinfo {author}
  {\bibfnamefont {A.}~\bibnamefont {Damascelli}},\ }\bibfield  {title}
  {\enquote {\bibinfo {title} {{${\mathrm{Na}}_{2}{\mathrm{IrO}}_{3}$ as a
  Novel Relativistic Mott Insulator with a 340-meV Gap}},}\ }\href {\doibase
  10.1103/PhysRevLett.109.266406} {\bibfield  {journal} {\bibinfo  {journal}
  {Phys. Rev. Lett.}\ }\textbf {\bibinfo {volume} {109}},\ \bibinfo {pages}
  {266406} (\bibinfo {year} {2012})}\BibitemShut {NoStop}%
\bibitem [{\citenamefont {Plumb}\ \emph {et~al.}(2014)\citenamefont {Plumb},
  \citenamefont {Clancy}, \citenamefont {Sandilands}, \citenamefont {Shankar},
  \citenamefont {Hu}, \citenamefont {Burch}, \citenamefont {Kee},\ and\
  \citenamefont {Kim}}]{PlumbPRB2014}%
  \BibitemOpen
  \bibfield  {author} {\bibinfo {author} {\bibfnamefont {K.~W.}\ \bibnamefont
  {Plumb}}, \bibinfo {author} {\bibfnamefont {J.~P.}\ \bibnamefont {Clancy}},
  \bibinfo {author} {\bibfnamefont {L.~J.}\ \bibnamefont {Sandilands}},
  \bibinfo {author} {\bibfnamefont {V.~Vijay}\ \bibnamefont {Shankar}},
  \bibinfo {author} {\bibfnamefont {Y.~F.}\ \bibnamefont {Hu}}, \bibinfo
  {author} {\bibfnamefont {K.~S.}\ \bibnamefont {Burch}}, \bibinfo {author}
  {\bibfnamefont {Hae-Young}\ \bibnamefont {Kee}}, \ and\ \bibinfo {author}
  {\bibfnamefont {Young-June}\ \bibnamefont {Kim}},\ }\bibfield  {title}
  {\enquote {\bibinfo {title}
  {{$\ensuremath{\alpha}\ensuremath{-}{\mathrm{RuCl}}_{3}$: A spin-orbit
  assisted Mott insulator on a honeycomb lattice}},}\ }\href {\doibase
  10.1103/PhysRevB.90.041112} {\bibfield  {journal} {\bibinfo  {journal} {Phys.
  Rev. B}\ }\textbf {\bibinfo {volume} {90}},\ \bibinfo {pages} {041112}
  (\bibinfo {year} {2014})}\BibitemShut {NoStop}%
\bibitem [{\citenamefont {Winter}\ \emph {et~al.}(2016)\citenamefont {Winter},
  \citenamefont {Li}, \citenamefont {Jeschke},\ and\ \citenamefont
  {Valent\'{\i}}}]{WinterPRB2016}%
  \BibitemOpen
  \bibfield  {author} {\bibinfo {author} {\bibfnamefont {Stephen~M.}\
  \bibnamefont {Winter}}, \bibinfo {author} {\bibfnamefont {Ying}\ \bibnamefont
  {Li}}, \bibinfo {author} {\bibfnamefont {Harald~O.}\ \bibnamefont {Jeschke}},
  \ and\ \bibinfo {author} {\bibfnamefont {Roser}\ \bibnamefont
  {Valent\'{\i}}},\ }\bibfield  {title} {\enquote {\bibinfo {title}
  {{Challenges in design of Kitaev materials: Magnetic interactions from
  competing energy scales}},}\ }\href {\doibase 10.1103/PhysRevB.93.214431}
  {\bibfield  {journal} {\bibinfo  {journal} {Phys. Rev. B}\ }\textbf {\bibinfo
  {volume} {93}},\ \bibinfo {pages} {214431} (\bibinfo {year}
  {2016})}\BibitemShut {NoStop}%
\bibitem [{\citenamefont {Lang}\ \emph {et~al.}(2016)\citenamefont {Lang},
  \citenamefont {Baker}, \citenamefont {Haghighirad}, \citenamefont {Li},
  \citenamefont {Prabhakaran}, \citenamefont {Valent\'{\i}},\ and\
  \citenamefont {Blundell}}]{LangPRB2016}%
  \BibitemOpen
  \bibfield  {author} {\bibinfo {author} {\bibfnamefont {F.}~\bibnamefont
  {Lang}}, \bibinfo {author} {\bibfnamefont {P.~J.}\ \bibnamefont {Baker}},
  \bibinfo {author} {\bibfnamefont {A.~A.}\ \bibnamefont {Haghighirad}},
  \bibinfo {author} {\bibfnamefont {Y.}~\bibnamefont {Li}}, \bibinfo {author}
  {\bibfnamefont {D.}~\bibnamefont {Prabhakaran}}, \bibinfo {author}
  {\bibfnamefont {R.}~\bibnamefont {Valent\'{\i}}}, \ and\ \bibinfo {author}
  {\bibfnamefont {S.~J.}\ \bibnamefont {Blundell}},\ }\bibfield  {title}
  {\enquote {\bibinfo {title} {{Unconventional magnetism on a honeycomb lattice
  in $\ensuremath{\alpha}\ensuremath{-}{\mathrm{RuCl}}_{3}$ studied by muon
  spin rotation}},}\ }\href {\doibase 10.1103/PhysRevB.94.020407} {\bibfield
  {journal} {\bibinfo  {journal} {Phys. Rev. B}\ }\textbf {\bibinfo {volume}
  {94}},\ \bibinfo {pages} {020407} (\bibinfo {year} {2016})}\BibitemShut
  {NoStop}%
\bibitem [{\citenamefont {Zhou}\ \emph {et~al.}(2017)\citenamefont {Zhou},
  \citenamefont {Kanoda},\ and\ \citenamefont {Ng}}]{ZhouRMP2017}%
  \BibitemOpen
  \bibfield  {author} {\bibinfo {author} {\bibfnamefont {Yi}~\bibnamefont
  {Zhou}}, \bibinfo {author} {\bibfnamefont {Kazushi}\ \bibnamefont {Kanoda}},
  \ and\ \bibinfo {author} {\bibfnamefont {Tai-Kai}\ \bibnamefont {Ng}},\
  }\bibfield  {title} {\enquote {\bibinfo {title} {{Quantum spin liquid
  states}},}\ }\href {\doibase 10.1103/RevModPhys.89.025003} {\bibfield
  {journal} {\bibinfo  {journal} {Rev. Mod. Phys.}\ }\textbf {\bibinfo {volume}
  {89}},\ \bibinfo {pages} {025003} (\bibinfo {year} {2017})}\BibitemShut
  {NoStop}%
\bibitem [{\citenamefont {Khaliullin}(2013)}]{KhaliullinPRL2013}%
  \BibitemOpen
  \bibfield  {author} {\bibinfo {author} {\bibfnamefont {Giniyat}\ \bibnamefont
  {Khaliullin}},\ }\bibfield  {title} {\enquote {\bibinfo {title} {{Excitonic
  Magnetism in Van Vleck--type ${d}^{4}$ Mott Insulators}},}\ }\href {\doibase
  10.1103/PhysRevLett.111.197201} {\bibfield  {journal} {\bibinfo  {journal}
  {Phys. Rev. Lett.}\ }\textbf {\bibinfo {volume} {111}},\ \bibinfo {pages}
  {197201} (\bibinfo {year} {2013})}\BibitemShut {NoStop}%
\bibitem [{\citenamefont {Sato}\ \emph {et~al.}(2019)\citenamefont {Sato},
  \citenamefont {Shirakawa},\ and\ \citenamefont {Yunoki}}]{SatoPRB2019}%
  \BibitemOpen
  \bibfield  {author} {\bibinfo {author} {\bibfnamefont {Toshihiro}\
  \bibnamefont {Sato}}, \bibinfo {author} {\bibfnamefont {Tomonori}\
  \bibnamefont {Shirakawa}}, \ and\ \bibinfo {author} {\bibfnamefont {Seiji}\
  \bibnamefont {Yunoki}},\ }\bibfield  {title} {\enquote {\bibinfo {title}
  {{Spin-orbital entangled excitonic insulator with quadrupole order}},}\
  }\href {\doibase 10.1103/PhysRevB.99.075117} {\bibfield  {journal} {\bibinfo
  {journal} {Phys. Rev. B}\ }\textbf {\bibinfo {volume} {99}},\ \bibinfo
  {pages} {075117} (\bibinfo {year} {2019})}\BibitemShut {NoStop}%
\bibitem [{\citenamefont {Chen}\ \emph {et~al.}(2010)\citenamefont {Chen},
  \citenamefont {Pereira},\ and\ \citenamefont {Balents}}]{ChenPRB2010}%
  \BibitemOpen
  \bibfield  {author} {\bibinfo {author} {\bibfnamefont {Gang}\ \bibnamefont
  {Chen}}, \bibinfo {author} {\bibfnamefont {Rodrigo}\ \bibnamefont {Pereira}},
  \ and\ \bibinfo {author} {\bibfnamefont {Leon}\ \bibnamefont {Balents}},\
  }\bibfield  {title} {\enquote {\bibinfo {title} {{Exotic phases induced by
  strong spin-orbit coupling in ordered double perovskites}},}\ }\href
  {\doibase 10.1103/PhysRevB.82.174440} {\bibfield  {journal} {\bibinfo
  {journal} {Phys. Rev. B}\ }\textbf {\bibinfo {volume} {82}},\ \bibinfo
  {pages} {174440} (\bibinfo {year} {2010})}\BibitemShut {NoStop}%
\bibitem [{\citenamefont {Chen}\ and\ \citenamefont
  {Balents}(2011)}]{ChenPRB2011}%
  \BibitemOpen
  \bibfield  {author} {\bibinfo {author} {\bibfnamefont {Gang}\ \bibnamefont
  {Chen}}\ and\ \bibinfo {author} {\bibfnamefont {Leon}\ \bibnamefont
  {Balents}},\ }\bibfield  {title} {\enquote {\bibinfo {title} {{Spin-orbit
  coupling in ${d}^{2}$ ordered double perovskites}},}\ }\href {\doibase
  10.1103/PhysRevB.84.094420} {\bibfield  {journal} {\bibinfo  {journal} {Phys.
  Rev. B}\ }\textbf {\bibinfo {volume} {84}},\ \bibinfo {pages} {094420}
  (\bibinfo {year} {2011})}\BibitemShut {NoStop}%
\bibitem [{\citenamefont {Meetei}\ \emph {et~al.}(2015)\citenamefont {Meetei},
  \citenamefont {Cole}, \citenamefont {Randeria},\ and\ \citenamefont
  {Trivedi}}]{MeeteiPRB2015}%
  \BibitemOpen
  \bibfield  {author} {\bibinfo {author} {\bibfnamefont {O.~Nganba}\
  \bibnamefont {Meetei}}, \bibinfo {author} {\bibfnamefont {William~S.}\
  \bibnamefont {Cole}}, \bibinfo {author} {\bibfnamefont {Mohit}\ \bibnamefont
  {Randeria}}, \ and\ \bibinfo {author} {\bibfnamefont {Nandini}\ \bibnamefont
  {Trivedi}},\ }\bibfield  {title} {\enquote {\bibinfo {title} {{Novel magnetic
  state in ${d}^{4}$ Mott insulators}},}\ }\href {\doibase
  10.1103/PhysRevB.91.054412} {\bibfield  {journal} {\bibinfo  {journal} {Phys.
  Rev. B}\ }\textbf {\bibinfo {volume} {91}},\ \bibinfo {pages} {054412}
  (\bibinfo {year} {2015})}\BibitemShut {NoStop}%
\bibitem [{\citenamefont {de' Medici}(2011)}]{MediciPRB2011}%
  \BibitemOpen
  \bibfield  {author} {\bibinfo {author} {\bibfnamefont {Luca}\ \bibnamefont
  {de' Medici}},\ }\bibfield  {title} {\enquote {\bibinfo {title} {Hund's
  coupling and its key role in tuning multiorbital correlations},}\ }\href
  {\doibase 10.1103/PhysRevB.83.205112} {\bibfield  {journal} {\bibinfo
  {journal} {Phys. Rev. B}\ }\textbf {\bibinfo {volume} {83}},\ \bibinfo
  {pages} {205112} (\bibinfo {year} {2011})}\BibitemShut {NoStop}%
\bibitem [{\citenamefont {Georges}\ \emph {et~al.}(2013)\citenamefont
  {Georges}, \citenamefont {Medici},\ and\ \citenamefont
  {Mravlje}}]{GeorgesARCMP2013}%
  \BibitemOpen
  \bibfield  {author} {\bibinfo {author} {\bibfnamefont {Antoine}\ \bibnamefont
  {Georges}}, \bibinfo {author} {\bibfnamefont {Luca~de'}\ \bibnamefont
  {Medici}}, \ and\ \bibinfo {author} {\bibfnamefont {Jernej}\ \bibnamefont
  {Mravlje}},\ }\bibfield  {title} {\enquote {\bibinfo {title} {{Strong
  Correlations from Hund's Coupling}},}\ }\href {\doibase
  10.1146/annurev-conmatphys-020911-125045} {\bibfield  {journal} {\bibinfo
  {journal} {Annu. Rev. Condens. Matter Phys.}\ }\textbf {\bibinfo {volume}
  {4}},\ \bibinfo {pages} {137} (\bibinfo {year} {2013})}\BibitemShut {NoStop}%
\bibitem [{\citenamefont {Gorelov}\ \emph {et~al.}(2010)\citenamefont
  {Gorelov}, \citenamefont {Karolak}, \citenamefont {Wehling}, \citenamefont
  {Lechermann}, \citenamefont {Lichtenstein},\ and\ \citenamefont
  {Pavarini}}]{GorelovPRL2010}%
  \BibitemOpen
  \bibfield  {author} {\bibinfo {author} {\bibfnamefont {E.}~\bibnamefont
  {Gorelov}}, \bibinfo {author} {\bibfnamefont {M.}~\bibnamefont {Karolak}},
  \bibinfo {author} {\bibfnamefont {T.~O.}\ \bibnamefont {Wehling}}, \bibinfo
  {author} {\bibfnamefont {F.}~\bibnamefont {Lechermann}}, \bibinfo {author}
  {\bibfnamefont {A.~I.}\ \bibnamefont {Lichtenstein}}, \ and\ \bibinfo
  {author} {\bibfnamefont {E.}~\bibnamefont {Pavarini}},\ }\bibfield  {title}
  {\enquote {\bibinfo {title} {{Nature of the Mott Transition in
  ${\mathrm{Ca}}_{2}{\mathrm{RuO}}_{4}$}},}\ }\href {\doibase
  10.1103/PhysRevLett.104.226401} {\bibfield  {journal} {\bibinfo  {journal}
  {Phys. Rev. Lett.}\ }\textbf {\bibinfo {volume} {104}},\ \bibinfo {pages}
  {226401} (\bibinfo {year} {2010})}\BibitemShut {NoStop}%
\bibitem [{\citenamefont {Mravlje}\ \emph {et~al.}(2011)\citenamefont
  {Mravlje}, \citenamefont {Aichhorn}, \citenamefont {Miyake}, \citenamefont
  {Haule}, \citenamefont {Kotliar},\ and\ \citenamefont
  {Georges}}]{MravljePRL2011}%
  \BibitemOpen
  \bibfield  {author} {\bibinfo {author} {\bibfnamefont {Jernej}\ \bibnamefont
  {Mravlje}}, \bibinfo {author} {\bibfnamefont {Markus}\ \bibnamefont
  {Aichhorn}}, \bibinfo {author} {\bibfnamefont {Takashi}\ \bibnamefont
  {Miyake}}, \bibinfo {author} {\bibfnamefont {Kristjan}\ \bibnamefont
  {Haule}}, \bibinfo {author} {\bibfnamefont {Gabriel}\ \bibnamefont
  {Kotliar}}, \ and\ \bibinfo {author} {\bibfnamefont {Antoine}\ \bibnamefont
  {Georges}},\ }\bibfield  {title} {\enquote {\bibinfo {title}
  {{Coherence-Incoherence Crossover and the Mass-Renormalization Puzzles in
  ${\mathrm{Sr}}_{2}{\mathrm{RuO}}_{4}$}},}\ }\href {\doibase
  10.1103/PhysRevLett.106.096401} {\bibfield  {journal} {\bibinfo  {journal}
  {Phys. Rev. Lett.}\ }\textbf {\bibinfo {volume} {106}},\ \bibinfo {pages}
  {096401} (\bibinfo {year} {2011})}\BibitemShut {NoStop}%
\bibitem [{\citenamefont {Stricker}\ \emph {et~al.}(2014)\citenamefont
  {Stricker}, \citenamefont {Mravlje}, \citenamefont {Berthod}, \citenamefont
  {Fittipaldi}, \citenamefont {Vecchione}, \citenamefont {Georges},\ and\
  \citenamefont {van~der Marel}}]{StrickerPRL2014}%
  \BibitemOpen
  \bibfield  {author} {\bibinfo {author} {\bibfnamefont {D.}~\bibnamefont
  {Stricker}}, \bibinfo {author} {\bibfnamefont {J.}~\bibnamefont {Mravlje}},
  \bibinfo {author} {\bibfnamefont {C.}~\bibnamefont {Berthod}}, \bibinfo
  {author} {\bibfnamefont {R.}~\bibnamefont {Fittipaldi}}, \bibinfo {author}
  {\bibfnamefont {A.}~\bibnamefont {Vecchione}}, \bibinfo {author}
  {\bibfnamefont {A.}~\bibnamefont {Georges}}, \ and\ \bibinfo {author}
  {\bibfnamefont {D.}~\bibnamefont {van~der Marel}},\ }\bibfield  {title}
  {\enquote {\bibinfo {title} {{Optical Response of
  ${\mathrm{Sr}}_{2}{\mathrm{RuO}}_{4}$ Reveals Universal Fermi-Liquid Scaling
  and Quasiparticles Beyond Landau Theory}},}\ }\href {\doibase
  10.1103/PhysRevLett.113.087404} {\bibfield  {journal} {\bibinfo  {journal}
  {Phys. Rev. Lett.}\ }\textbf {\bibinfo {volume} {113}},\ \bibinfo {pages}
  {087404} (\bibinfo {year} {2014})}\BibitemShut {NoStop}%
\bibitem [{\citenamefont {Dang}\ \emph
  {et~al.}(2015{\natexlab{a}})\citenamefont {Dang}, \citenamefont {Mravlje},
  \citenamefont {Georges},\ and\ \citenamefont {Millis}}]{DangPRL2015}%
  \BibitemOpen
  \bibfield  {author} {\bibinfo {author} {\bibfnamefont {Hung~T.}\ \bibnamefont
  {Dang}}, \bibinfo {author} {\bibfnamefont {Jernej}\ \bibnamefont {Mravlje}},
  \bibinfo {author} {\bibfnamefont {Antoine}\ \bibnamefont {Georges}}, \ and\
  \bibinfo {author} {\bibfnamefont {Andrew~J.}\ \bibnamefont {Millis}},\
  }\bibfield  {title} {\enquote {\bibinfo {title} {{Band Structure and
  Terahertz Optical Conductivity of Transition Metal Oxides: Theory and
  Application to ${\mathrm{CaRuO}}_{3}$}},}\ }\href {\doibase
  10.1103/PhysRevLett.115.107003} {\bibfield  {journal} {\bibinfo  {journal}
  {Phys. Rev. Lett.}\ }\textbf {\bibinfo {volume} {115}},\ \bibinfo {pages}
  {107003} (\bibinfo {year} {2015}{\natexlab{a}})}\BibitemShut {NoStop}%
\bibitem [{\citenamefont {Dang}\ \emph
  {et~al.}(2015{\natexlab{b}})\citenamefont {Dang}, \citenamefont {Mravlje},
  \citenamefont {Georges},\ and\ \citenamefont {Millis}}]{DangPRB2015}%
  \BibitemOpen
  \bibfield  {author} {\bibinfo {author} {\bibfnamefont {Hung~T.}\ \bibnamefont
  {Dang}}, \bibinfo {author} {\bibfnamefont {Jernej}\ \bibnamefont {Mravlje}},
  \bibinfo {author} {\bibfnamefont {Antoine}\ \bibnamefont {Georges}}, \ and\
  \bibinfo {author} {\bibfnamefont {Andrew~J.}\ \bibnamefont {Millis}},\
  }\bibfield  {title} {\enquote {\bibinfo {title} {{Electronic correlations,
  magnetism, and Hund's rule coupling in the ruthenium perovskites
  ${\text{SrRuO}}_{3}$ and ${\text{CaRuO}}_{3}$}},}\ }\href {\doibase
  10.1103/PhysRevB.91.195149} {\bibfield  {journal} {\bibinfo  {journal} {Phys.
  Rev. B}\ }\textbf {\bibinfo {volume} {91}},\ \bibinfo {pages} {195149}
  (\bibinfo {year} {2015}{\natexlab{b}})}\BibitemShut {NoStop}%
\bibitem [{\citenamefont {Sutter}\ \emph {et~al.}(2019)\citenamefont {Sutter},
  \citenamefont {Kim}, \citenamefont {Matt}, \citenamefont {Horio},
  \citenamefont {Fittipaldi}, \citenamefont {Vecchione}, \citenamefont
  {Granata}, \citenamefont {Hauser}, \citenamefont {Sassa}, \citenamefont
  {Gatti}, \citenamefont {Grioni}, \citenamefont {Hoesch}, \citenamefont {Kim},
  \citenamefont {Rienks}, \citenamefont {Plumb}, \citenamefont {Shi},
  \citenamefont {Neupert}, \citenamefont {Georges},\ and\ \citenamefont
  {Chang}}]{SutterPRB2019}%
  \BibitemOpen
  \bibfield  {author} {\bibinfo {author} {\bibfnamefont {D.}~\bibnamefont
  {Sutter}}, \bibinfo {author} {\bibfnamefont {M.}~\bibnamefont {Kim}},
  \bibinfo {author} {\bibfnamefont {C.~E.}\ \bibnamefont {Matt}}, \bibinfo
  {author} {\bibfnamefont {M.}~\bibnamefont {Horio}}, \bibinfo {author}
  {\bibfnamefont {R.}~\bibnamefont {Fittipaldi}}, \bibinfo {author}
  {\bibfnamefont {A.}~\bibnamefont {Vecchione}}, \bibinfo {author}
  {\bibfnamefont {V.}~\bibnamefont {Granata}}, \bibinfo {author} {\bibfnamefont
  {K.}~\bibnamefont {Hauser}}, \bibinfo {author} {\bibfnamefont
  {Y.}~\bibnamefont {Sassa}}, \bibinfo {author} {\bibfnamefont
  {G.}~\bibnamefont {Gatti}}, \bibinfo {author} {\bibfnamefont
  {M.}~\bibnamefont {Grioni}}, \bibinfo {author} {\bibfnamefont
  {M.}~\bibnamefont {Hoesch}}, \bibinfo {author} {\bibfnamefont {T.~K.}\
  \bibnamefont {Kim}}, \bibinfo {author} {\bibfnamefont {E.}~\bibnamefont
  {Rienks}}, \bibinfo {author} {\bibfnamefont {N.~C.}\ \bibnamefont {Plumb}},
  \bibinfo {author} {\bibfnamefont {M.}~\bibnamefont {Shi}}, \bibinfo {author}
  {\bibfnamefont {T.}~\bibnamefont {Neupert}}, \bibinfo {author} {\bibfnamefont
  {A.}~\bibnamefont {Georges}}, \ and\ \bibinfo {author} {\bibfnamefont
  {J.}~\bibnamefont {Chang}},\ }\bibfield  {title} {\enquote {\bibinfo {title}
  {{Orbitally selective breakdown of Fermi liquid quasiparticles in
  ${\mathrm{Ca}}_{1.8}{\mathrm{Sr}}_{0.2}{\mathrm{RuO}}_{4}$}},}\ }\href
  {\doibase 10.1103/PhysRevB.99.121115} {\bibfield  {journal} {\bibinfo
  {journal} {Phys. Rev. B}\ }\textbf {\bibinfo {volume} {99}},\ \bibinfo
  {pages} {121115} (\bibinfo {year} {2019})}\BibitemShut {NoStop}%
\bibitem [{\citenamefont {Kugler}\ \emph {et~al.}(2020)\citenamefont {Kugler},
  \citenamefont {Zingl}, \citenamefont {Strand}, \citenamefont {Lee},
  \citenamefont {von Delft},\ and\ \citenamefont {Georges}}]{KuglerPRL2020}%
  \BibitemOpen
  \bibfield  {author} {\bibinfo {author} {\bibfnamefont {Fabian~B.}\
  \bibnamefont {Kugler}}, \bibinfo {author} {\bibfnamefont {Manuel}\
  \bibnamefont {Zingl}}, \bibinfo {author} {\bibfnamefont {Hugo U.~R.}\
  \bibnamefont {Strand}}, \bibinfo {author} {\bibfnamefont {Seung-Sup~B.}\
  \bibnamefont {Lee}}, \bibinfo {author} {\bibfnamefont {Jan}\ \bibnamefont
  {von Delft}}, \ and\ \bibinfo {author} {\bibfnamefont {Antoine}\ \bibnamefont
  {Georges}},\ }\bibfield  {title} {\enquote {\bibinfo {title} {{Strongly
  Correlated Materials from a Numerical Renormalization Group Perspective: How
  the Fermi-Liquid State of ${\mathrm{Sr}}_{2}{\mathrm{RuO}}_{4}$ Emerges}},}\
  }\href {\doibase 10.1103/PhysRevLett.124.016401} {\bibfield  {journal}
  {\bibinfo  {journal} {Phys. Rev. Lett.}\ }\textbf {\bibinfo {volume} {124}},\
  \bibinfo {pages} {016401} (\bibinfo {year} {2020})}\BibitemShut {NoStop}%
\bibitem [{\citenamefont {Haule}\ and\ \citenamefont
  {Kotliar}(2009)}]{HauleNJP2009}%
  \BibitemOpen
  \bibfield  {author} {\bibinfo {author} {\bibfnamefont {K.}~\bibnamefont
  {Haule}}\ and\ \bibinfo {author} {\bibfnamefont {G.}~\bibnamefont
  {Kotliar}},\ }\bibfield  {title} {\enquote {\bibinfo {title}
  {{Coherence-incoherence crossover in the normal state of iron oxypnictides
  and importance of Hund's rule coupling}},}\ }\href {\doibase
  10.1088/1367-2630/11/2/025021} {\bibfield  {journal} {\bibinfo  {journal}
  {New J. Phys.}\ }\textbf {\bibinfo {volume} {11}},\ \bibinfo {pages} {025021}
  (\bibinfo {year} {2009})}\BibitemShut {NoStop}%
\bibitem [{\citenamefont {Yin}\ \emph {et~al.}(2011)\citenamefont {Yin},
  \citenamefont {Haule},\ and\ \citenamefont {Kotliar}}]{YinNatPhys2011}%
  \BibitemOpen
  \bibfield  {author} {\bibinfo {author} {\bibfnamefont {Z.~P.}\ \bibnamefont
  {Yin}}, \bibinfo {author} {\bibfnamefont {K.}~\bibnamefont {Haule}}, \ and\
  \bibinfo {author} {\bibfnamefont {G.}~\bibnamefont {Kotliar}},\ }\bibfield
  {title} {\enquote {\bibinfo {title} {{Magnetism and charge dynamics in iron
  pnictides}},}\ }\href {\doibase 10.1038/NPHYS1923} {\bibfield  {journal}
  {\bibinfo  {journal} {Nature Physics}\ }\textbf {\bibinfo {volume} {7}},\
  \bibinfo {pages} {294} (\bibinfo {year} {2011})}\BibitemShut {NoStop}%
\bibitem [{\citenamefont {Lanat\`a}\ \emph {et~al.}(2013)\citenamefont
  {Lanat\`a}, \citenamefont {Strand}, \citenamefont {Giovannetti},
  \citenamefont {Hellsing}, \citenamefont {de' Medici},\ and\ \citenamefont
  {Capone}}]{NicolaPRB2013}%
  \BibitemOpen
  \bibfield  {author} {\bibinfo {author} {\bibfnamefont {Nicola}\ \bibnamefont
  {Lanat\`a}}, \bibinfo {author} {\bibfnamefont {Hugo U.~R.}\ \bibnamefont
  {Strand}}, \bibinfo {author} {\bibfnamefont {Gianluca}\ \bibnamefont
  {Giovannetti}}, \bibinfo {author} {\bibfnamefont {Bo}~\bibnamefont
  {Hellsing}}, \bibinfo {author} {\bibfnamefont {Luca}\ \bibnamefont {de'
  Medici}}, \ and\ \bibinfo {author} {\bibfnamefont {Massimo}\ \bibnamefont
  {Capone}},\ }\bibfield  {title} {\enquote {\bibinfo {title} {{Orbital
  selectivity in Hund's metals: The iron chalcogenides}},}\ }\href {\doibase
  10.1103/PhysRevB.87.045122} {\bibfield  {journal} {\bibinfo  {journal} {Phys.
  Rev. B}\ }\textbf {\bibinfo {volume} {87}},\ \bibinfo {pages} {045122}
  (\bibinfo {year} {2013})}\BibitemShut {NoStop}%
\bibitem [{\citenamefont {Yi}\ \emph {et~al.}(2013)\citenamefont {Yi},
  \citenamefont {Lu}, \citenamefont {Yu}, \citenamefont {Riggs}, \citenamefont
  {Chu}, \citenamefont {Lv}, \citenamefont {Liu}, \citenamefont {Lu},
  \citenamefont {Cui}, \citenamefont {Hashimoto}, \citenamefont {Mo},
  \citenamefont {Hussain}, \citenamefont {Chu}, \citenamefont {Fisher},
  \citenamefont {Si},\ and\ \citenamefont {Shen}}]{YiPRL2013}%
  \BibitemOpen
  \bibfield  {author} {\bibinfo {author} {\bibfnamefont {M.}~\bibnamefont
  {Yi}}, \bibinfo {author} {\bibfnamefont {D.~H.}\ \bibnamefont {Lu}}, \bibinfo
  {author} {\bibfnamefont {R.}~\bibnamefont {Yu}}, \bibinfo {author}
  {\bibfnamefont {S.~C.}\ \bibnamefont {Riggs}}, \bibinfo {author}
  {\bibfnamefont {J.-H.}\ \bibnamefont {Chu}}, \bibinfo {author} {\bibfnamefont
  {B.}~\bibnamefont {Lv}}, \bibinfo {author} {\bibfnamefont {Z.~K.}\
  \bibnamefont {Liu}}, \bibinfo {author} {\bibfnamefont {M.}~\bibnamefont
  {Lu}}, \bibinfo {author} {\bibfnamefont {Y.-T.}\ \bibnamefont {Cui}},
  \bibinfo {author} {\bibfnamefont {M.}~\bibnamefont {Hashimoto}}, \bibinfo
  {author} {\bibfnamefont {S.-K.}\ \bibnamefont {Mo}}, \bibinfo {author}
  {\bibfnamefont {Z.}~\bibnamefont {Hussain}}, \bibinfo {author} {\bibfnamefont
  {C.~W.}\ \bibnamefont {Chu}}, \bibinfo {author} {\bibfnamefont {I.~R.}\
  \bibnamefont {Fisher}}, \bibinfo {author} {\bibfnamefont {Q.}~\bibnamefont
  {Si}}, \ and\ \bibinfo {author} {\bibfnamefont {Z.-X.}\ \bibnamefont
  {Shen}},\ }\bibfield  {title} {\enquote {\bibinfo {title} {{Observation of
  Temperature-Induced Crossover to an Orbital-Selective Mott Phase in
  ${\mathrm{A}}_{x}{\mathrm{Fe}}_{2\mathrm{\text{\ensuremath{-}}}y}{\mathrm{Se}}_{2}$
  ($A\mathbf{=}\mathrm{K}$, Rb) Superconductors}},}\ }\href {\doibase
  10.1103/PhysRevLett.110.067003} {\bibfield  {journal} {\bibinfo  {journal}
  {Phys. Rev. Lett.}\ }\textbf {\bibinfo {volume} {110}},\ \bibinfo {pages}
  {067003} (\bibinfo {year} {2013})}\BibitemShut {NoStop}%
\bibitem [{\citenamefont {Sprau}\ \emph {et~al.}(2017)\citenamefont {Sprau},
  \citenamefont {Kostin}, \citenamefont {Kreisel}, \citenamefont {B{\"o}hmer},
  \citenamefont {Taufour}, \citenamefont {Canfield}, \citenamefont {Mukherjee},
  \citenamefont {Hirschfeld}, \citenamefont {Andersen},\ and\ \citenamefont
  {Davis}}]{SprauScience2017}%
  \BibitemOpen
  \bibfield  {author} {\bibinfo {author} {\bibfnamefont {P.~O.}\ \bibnamefont
  {Sprau}}, \bibinfo {author} {\bibfnamefont {A.}~\bibnamefont {Kostin}},
  \bibinfo {author} {\bibfnamefont {A.}~\bibnamefont {Kreisel}}, \bibinfo
  {author} {\bibfnamefont {A.~E.}\ \bibnamefont {B{\"o}hmer}}, \bibinfo
  {author} {\bibfnamefont {V.}~\bibnamefont {Taufour}}, \bibinfo {author}
  {\bibfnamefont {P.~C.}\ \bibnamefont {Canfield}}, \bibinfo {author}
  {\bibfnamefont {S.}~\bibnamefont {Mukherjee}}, \bibinfo {author}
  {\bibfnamefont {P.~J.}\ \bibnamefont {Hirschfeld}}, \bibinfo {author}
  {\bibfnamefont {B.~M.}\ \bibnamefont {Andersen}}, \ and\ \bibinfo {author}
  {\bibfnamefont {J.~C.~S{\'e}amus}\ \bibnamefont {Davis}},\ }\bibfield
  {title} {\enquote {\bibinfo {title} {{Discovery of orbital-selective Cooper
  pairing in FeSe}},}\ }\href {\doibase 10.1126/science.aal1575} {\bibfield
  {journal} {\bibinfo  {journal} {Science}\ }\textbf {\bibinfo {volume}
  {357}},\ \bibinfo {pages} {75} (\bibinfo {year} {2017})}\BibitemShut
  {NoStop}%
\bibitem [{\citenamefont {Koga}\ \emph {et~al.}(2004)\citenamefont {Koga},
  \citenamefont {Kawakami}, \citenamefont {Rice},\ and\ \citenamefont
  {Sigrist}}]{KogaPRL2004}%
  \BibitemOpen
  \bibfield  {author} {\bibinfo {author} {\bibfnamefont {Akihisa}\ \bibnamefont
  {Koga}}, \bibinfo {author} {\bibfnamefont {Norio}\ \bibnamefont {Kawakami}},
  \bibinfo {author} {\bibfnamefont {T.~M.}\ \bibnamefont {Rice}}, \ and\
  \bibinfo {author} {\bibfnamefont {Manfred}\ \bibnamefont {Sigrist}},\
  }\bibfield  {title} {\enquote {\bibinfo {title} {{Orbital-Selective Mott
  Transitions in the Degenerate Hubbard Model}},}\ }\href {\doibase
  10.1103/PhysRevLett.92.216402} {\bibfield  {journal} {\bibinfo  {journal}
  {Phys. Rev. Lett.}\ }\textbf {\bibinfo {volume} {92}},\ \bibinfo {pages}
  {216402} (\bibinfo {year} {2004})}\BibitemShut {NoStop}%
\bibitem [{\citenamefont {Werner}\ and\ \citenamefont
  {Millis}(2007)}]{WernerPRL2007}%
  \BibitemOpen
  \bibfield  {author} {\bibinfo {author} {\bibfnamefont {Philipp}\ \bibnamefont
  {Werner}}\ and\ \bibinfo {author} {\bibfnamefont {Andrew~J.}\ \bibnamefont
  {Millis}},\ }\bibfield  {title} {\enquote {\bibinfo {title} {{High-Spin to
  Low-Spin and Orbital Polarization Transitions in Multiorbital Mott
  Systems}},}\ }\href {\doibase 10.1103/PhysRevLett.99.126405} {\bibfield
  {journal} {\bibinfo  {journal} {Phys. Rev. Lett.}\ }\textbf {\bibinfo
  {volume} {99}},\ \bibinfo {pages} {126405} (\bibinfo {year}
  {2007})}\BibitemShut {NoStop}%
\bibitem [{\citenamefont {de' Medici}\ \emph {et~al.}(2009)\citenamefont {de'
  Medici}, \citenamefont {Hassan}, \citenamefont {Capone},\ and\ \citenamefont
  {Dai}}]{MediciPRL2009}%
  \BibitemOpen
  \bibfield  {author} {\bibinfo {author} {\bibfnamefont {Luca}\ \bibnamefont
  {de' Medici}}, \bibinfo {author} {\bibfnamefont {S.~R.}\ \bibnamefont
  {Hassan}}, \bibinfo {author} {\bibfnamefont {Massimo}\ \bibnamefont
  {Capone}}, \ and\ \bibinfo {author} {\bibfnamefont {Xi}~\bibnamefont {Dai}},\
  }\bibfield  {title} {\enquote {\bibinfo {title} {{Orbital-Selective Mott
  Transition out of Band Degeneracy Lifting}},}\ }\href {\doibase
  10.1103/PhysRevLett.102.126401} {\bibfield  {journal} {\bibinfo  {journal}
  {Phys. Rev. Lett.}\ }\textbf {\bibinfo {volume} {102}},\ \bibinfo {pages}
  {126401} (\bibinfo {year} {2009})}\BibitemShut {NoStop}%
\bibitem [{\citenamefont {Lee}\ \emph {et~al.}(2011)\citenamefont {Lee},
  \citenamefont {Zhang}, \citenamefont {Jeschke},\ and\ \citenamefont
  {Valent\'{\i}}}]{LeePRB2011}%
  \BibitemOpen
  \bibfield  {author} {\bibinfo {author} {\bibfnamefont {Hunpyo}\ \bibnamefont
  {Lee}}, \bibinfo {author} {\bibfnamefont {Yu-Zhong}\ \bibnamefont {Zhang}},
  \bibinfo {author} {\bibfnamefont {Harald~O.}\ \bibnamefont {Jeschke}}, \ and\
  \bibinfo {author} {\bibfnamefont {Roser}\ \bibnamefont {Valent\'{\i}}},\
  }\bibfield  {title} {\enquote {\bibinfo {title} {{Orbital-selective phase
  transition induced by different magnetic states: A dynamical cluster
  approximation study}},}\ }\href {\doibase 10.1103/PhysRevB.84.020401}
  {\bibfield  {journal} {\bibinfo  {journal} {Phys. Rev. B}\ }\textbf {\bibinfo
  {volume} {84}},\ \bibinfo {pages} {020401} (\bibinfo {year}
  {2011})}\BibitemShut {NoStop}%
\bibitem [{\citenamefont {Song}\ \emph {et~al.}(2015)\citenamefont {Song},
  \citenamefont {Lee},\ and\ \citenamefont {Zhang}}]{SongNJP2015}%
  \BibitemOpen
  \bibfield  {author} {\bibinfo {author} {\bibfnamefont {Ze-Yi}\ \bibnamefont
  {Song}}, \bibinfo {author} {\bibfnamefont {Hunpyo}\ \bibnamefont {Lee}}, \
  and\ \bibinfo {author} {\bibfnamefont {Yu-Zhong}\ \bibnamefont {Zhang}},\
  }\bibfield  {title} {\enquote {\bibinfo {title} {{Possible origin of orbital
  selective Mott transitions in iron-based superconductors and
  Ca$_{2-x}$Sr$_x$RuO$_4$}},}\ }\href {\doibase 10.1088/1367-2630/17/3/033034}
  {\bibfield  {journal} {\bibinfo  {journal} {New J. Phys.}\ }\textbf {\bibinfo
  {volume} {17}},\ \bibinfo {pages} {033034} (\bibinfo {year}
  {2015})}\BibitemShut {NoStop}%
\bibitem [{\citenamefont {Anisimov}\ \emph {et~al.}(2002)\citenamefont
  {Anisimov}, \citenamefont {Nekrasov}, \citenamefont {Kondakov}, \citenamefont
  {Rice},\ and\ \citenamefont {Sigrist}}]{AnisimovEPJB2002}%
  \BibitemOpen
  \bibfield  {author} {\bibinfo {author} {\bibfnamefont {V.~I.}\ \bibnamefont
  {Anisimov}}, \bibinfo {author} {\bibfnamefont {I.~A.}\ \bibnamefont
  {Nekrasov}}, \bibinfo {author} {\bibfnamefont {D.~E.}\ \bibnamefont
  {Kondakov}}, \bibinfo {author} {\bibfnamefont {T.~M.}\ \bibnamefont {Rice}},
  \ and\ \bibinfo {author} {\bibfnamefont {M.}~\bibnamefont {Sigrist}},\
  }\bibfield  {title} {\enquote {\bibinfo {title} {{Orbital-selective
  Mott-insulator transition in Ca$_{2-x}$Sr$_x$RuO$_4$}},}\ }\href {\doibase
  10.1140/epjb/e20020021} {\bibfield  {journal} {\bibinfo  {journal} {Eur.
  Phys. J. B}\ }\textbf {\bibinfo {volume} {25}},\ \bibinfo {pages} {191}
  (\bibinfo {year} {2002})}\BibitemShut {NoStop}%
\bibitem [{\citenamefont {Shimoyamada}\ \emph {et~al.}(2009)\citenamefont
  {Shimoyamada}, \citenamefont {Ishizaka}, \citenamefont {Tsuda}, \citenamefont
  {Nakatsuji}, \citenamefont {Maeno},\ and\ \citenamefont
  {Shin}}]{ShimoyamadaPRL2009}%
  \BibitemOpen
  \bibfield  {author} {\bibinfo {author} {\bibfnamefont {A.}~\bibnamefont
  {Shimoyamada}}, \bibinfo {author} {\bibfnamefont {K.}~\bibnamefont
  {Ishizaka}}, \bibinfo {author} {\bibfnamefont {S.}~\bibnamefont {Tsuda}},
  \bibinfo {author} {\bibfnamefont {S.}~\bibnamefont {Nakatsuji}}, \bibinfo
  {author} {\bibfnamefont {Y.}~\bibnamefont {Maeno}}, \ and\ \bibinfo {author}
  {\bibfnamefont {S.}~\bibnamefont {Shin}},\ }\bibfield  {title} {\enquote
  {\bibinfo {title} {{Strong Mass Renormalization at a Local Momentum Space in
  Multiorbital ${\mathrm{Ca}}_{1.8}{\mathrm{Sr}}_{0.2}{\mathrm{RuO}}_{4}$}},}\
  }\href {\doibase 10.1103/PhysRevLett.102.086401} {\bibfield  {journal}
  {\bibinfo  {journal} {Phys. Rev. Lett.}\ }\textbf {\bibinfo {volume} {102}},\
  \bibinfo {pages} {086401} (\bibinfo {year} {2009})}\BibitemShut {NoStop}%
\bibitem [{\citenamefont {Neupane}\ \emph {et~al.}(2009)\citenamefont
  {Neupane}, \citenamefont {Richard}, \citenamefont {Pan}, \citenamefont {Xu},
  \citenamefont {Jin}, \citenamefont {Mandrus}, \citenamefont {Dai},
  \citenamefont {Fang}, \citenamefont {Wang},\ and\ \citenamefont
  {Ding}}]{NeupanePRL2009}%
  \BibitemOpen
  \bibfield  {author} {\bibinfo {author} {\bibfnamefont {M.}~\bibnamefont
  {Neupane}}, \bibinfo {author} {\bibfnamefont {P.}~\bibnamefont {Richard}},
  \bibinfo {author} {\bibfnamefont {Z.-H.}\ \bibnamefont {Pan}}, \bibinfo
  {author} {\bibfnamefont {Y.-M.}\ \bibnamefont {Xu}}, \bibinfo {author}
  {\bibfnamefont {R.}~\bibnamefont {Jin}}, \bibinfo {author} {\bibfnamefont
  {D.}~\bibnamefont {Mandrus}}, \bibinfo {author} {\bibfnamefont
  {X.}~\bibnamefont {Dai}}, \bibinfo {author} {\bibfnamefont {Z.}~\bibnamefont
  {Fang}}, \bibinfo {author} {\bibfnamefont {Z.}~\bibnamefont {Wang}}, \ and\
  \bibinfo {author} {\bibfnamefont {H.}~\bibnamefont {Ding}},\ }\bibfield
  {title} {\enquote {\bibinfo {title} {{Observation of a Novel Orbital
  Selective Mott Transition in
  ${\mathrm{Ca}}_{1.8}{\mathrm{Sr}}_{0.2}{\mathrm{RuO}}_{4}$}},}\ }\href
  {\doibase 10.1103/PhysRevLett.103.097001} {\bibfield  {journal} {\bibinfo
  {journal} {Phys. Rev. Lett.}\ }\textbf {\bibinfo {volume} {103}},\ \bibinfo
  {pages} {097001} (\bibinfo {year} {2009})}\BibitemShut {NoStop}%
\bibitem [{\citenamefont {Huang}\ \emph {et~al.}(2012)\citenamefont {Huang},
  \citenamefont {Du},\ and\ \citenamefont {Dai}}]{HuangPRB2012}%
  \BibitemOpen
  \bibfield  {author} {\bibinfo {author} {\bibfnamefont {Li}~\bibnamefont
  {Huang}}, \bibinfo {author} {\bibfnamefont {Liang}\ \bibnamefont {Du}}, \
  and\ \bibinfo {author} {\bibfnamefont {Xi}~\bibnamefont {Dai}},\ }\bibfield
  {title} {\enquote {\bibinfo {title} {{Complete phase diagram for three-band
  Hubbard model with orbital degeneracy lifted by crystal field splitting}},}\
  }\href {\doibase 10.1103/PhysRevB.86.035150} {\bibfield  {journal} {\bibinfo
  {journal} {Phys. Rev. B}\ }\textbf {\bibinfo {volume} {86}},\ \bibinfo
  {pages} {035150} (\bibinfo {year} {2012})}\BibitemShut {NoStop}%
\bibitem [{\citenamefont {Du}\ \emph {et~al.}(2013)\citenamefont {Du},
  \citenamefont {Huang},\ and\ \citenamefont {Dai}}]{DuEPJB2013}%
  \BibitemOpen
  \bibfield  {author} {\bibinfo {author} {\bibfnamefont {Liang}\ \bibnamefont
  {Du}}, \bibinfo {author} {\bibfnamefont {Li}~\bibnamefont {Huang}}, \ and\
  \bibinfo {author} {\bibfnamefont {Xi}~\bibnamefont {Dai}},\ }\bibfield
  {title} {\enquote {\bibinfo {title} {{Metal-insulator transition in
  three-band Hubbard model with strong spin-orbit interaction}},}\ }\href
  {\doibase 10.1140/epjb/e2013-31024-6} {\bibfield  {journal} {\bibinfo
  {journal} {Eur. Phys. J. B}\ }\textbf {\bibinfo {volume} {86}},\ \bibinfo
  {pages} {94} (\bibinfo {year} {2013})}\BibitemShut {NoStop}%
\bibitem [{\citenamefont {Kim}\ \emph {et~al.}(2017)\citenamefont {Kim},
  \citenamefont {Jeschke}, \citenamefont {Werner},\ and\ \citenamefont
  {Valent\'{\i}}}]{KimPRL2017}%
  \BibitemOpen
  \bibfield  {author} {\bibinfo {author} {\bibfnamefont {Aaram~J.}\
  \bibnamefont {Kim}}, \bibinfo {author} {\bibfnamefont {Harald~O.}\
  \bibnamefont {Jeschke}}, \bibinfo {author} {\bibfnamefont {Philipp}\
  \bibnamefont {Werner}}, \ and\ \bibinfo {author} {\bibfnamefont {Roser}\
  \bibnamefont {Valent\'{\i}}},\ }\bibfield  {title} {\enquote {\bibinfo
  {title} {{$\mathbf{J}$ Freezing and Hund's Rules in Spin-Orbit-Coupled
  Multiorbital Hubbard Models}},}\ }\href {\doibase
  10.1103/PhysRevLett.118.086401} {\bibfield  {journal} {\bibinfo  {journal}
  {Phys. Rev. Lett.}\ }\textbf {\bibinfo {volume} {118}},\ \bibinfo {pages}
  {086401} (\bibinfo {year} {2017})}\BibitemShut {NoStop}%
\bibitem [{\citenamefont {Triebl}\ \emph {et~al.}(2018)\citenamefont {Triebl},
  \citenamefont {Kraberger}, \citenamefont {Mravlje},\ and\ \citenamefont
  {Aichhorn}}]{TrieblPRB2018}%
  \BibitemOpen
  \bibfield  {author} {\bibinfo {author} {\bibfnamefont {Robert}\ \bibnamefont
  {Triebl}}, \bibinfo {author} {\bibfnamefont {Gernot~J.}\ \bibnamefont
  {Kraberger}}, \bibinfo {author} {\bibfnamefont {Jernej}\ \bibnamefont
  {Mravlje}}, \ and\ \bibinfo {author} {\bibfnamefont {Markus}\ \bibnamefont
  {Aichhorn}},\ }\bibfield  {title} {\enquote {\bibinfo {title} {{Spin-orbit
  coupling and correlations in three-orbital systems}},}\ }\href {\doibase
  10.1103/PhysRevB.98.205128} {\bibfield  {journal} {\bibinfo  {journal} {Phys.
  Rev. B}\ }\textbf {\bibinfo {volume} {98}},\ \bibinfo {pages} {205128}
  (\bibinfo {year} {2018})}\BibitemShut {NoStop}%
\bibitem [{\citenamefont {Piefke}\ and\ \citenamefont
  {Lechermann}(2018)}]{PiefkePRB2018}%
  \BibitemOpen
  \bibfield  {author} {\bibinfo {author} {\bibfnamefont {Christoph}\
  \bibnamefont {Piefke}}\ and\ \bibinfo {author} {\bibfnamefont {Frank}\
  \bibnamefont {Lechermann}},\ }\bibfield  {title} {\enquote {\bibinfo {title}
  {{Rigorous symmetry adaptation of multiorbital rotationally invariant
  slave-boson theory with application to Hund's rules physics}},}\ }\href
  {\doibase 10.1103/PhysRevB.97.125154} {\bibfield  {journal} {\bibinfo
  {journal} {Phys. Rev. B}\ }\textbf {\bibinfo {volume} {97}},\ \bibinfo
  {pages} {125154} (\bibinfo {year} {2018})}\BibitemShut {NoStop}%
\bibitem [{\citenamefont {Georges}\ \emph {et~al.}(1996)\citenamefont
  {Georges}, \citenamefont {Kotliar}, \citenamefont {Krauth},\ and\
  \citenamefont {Rozenberg}}]{GeorgesRMP1996}%
  \BibitemOpen
  \bibfield  {author} {\bibinfo {author} {\bibfnamefont {Antoine}\ \bibnamefont
  {Georges}}, \bibinfo {author} {\bibfnamefont {Gabriel}\ \bibnamefont
  {Kotliar}}, \bibinfo {author} {\bibfnamefont {Werner}\ \bibnamefont
  {Krauth}}, \ and\ \bibinfo {author} {\bibfnamefont {Marcelo~J.}\ \bibnamefont
  {Rozenberg}},\ }\bibfield  {title} {\enquote {\bibinfo {title} {{Dynamical
  mean-field theory of strongly correlated fermion systems and the limit of
  infinite dimensions}},}\ }\href {\doibase 10.1103/RevModPhys.68.13}
  {\bibfield  {journal} {\bibinfo  {journal} {Rev. Mod. Phys.}\ }\textbf
  {\bibinfo {volume} {68}},\ \bibinfo {pages} {13} (\bibinfo {year}
  {1996})}\BibitemShut {NoStop}%
\bibitem [{\citenamefont {Cooper}\ \emph {et~al.}(2009)\citenamefont {Cooper},
  \citenamefont {Wang}, \citenamefont {Vignolle}, \citenamefont {Lipscombe},
  \citenamefont {Hayden}, \citenamefont {Tanabe}, \citenamefont {Adachi},
  \citenamefont {Koike}, \citenamefont {Nohara}, \citenamefont {Takagi},
  \citenamefont {Proust},\ and\ \citenamefont {Hussey}}]{CooperScience2009}%
  \BibitemOpen
  \bibfield  {author} {\bibinfo {author} {\bibfnamefont {R.~A.}\ \bibnamefont
  {Cooper}}, \bibinfo {author} {\bibfnamefont {Y.}~\bibnamefont {Wang}},
  \bibinfo {author} {\bibfnamefont {B.}~\bibnamefont {Vignolle}}, \bibinfo
  {author} {\bibfnamefont {O.~J.}\ \bibnamefont {Lipscombe}}, \bibinfo {author}
  {\bibfnamefont {S.~M.}\ \bibnamefont {Hayden}}, \bibinfo {author}
  {\bibfnamefont {Y.}~\bibnamefont {Tanabe}}, \bibinfo {author} {\bibfnamefont
  {T.}~\bibnamefont {Adachi}}, \bibinfo {author} {\bibfnamefont
  {Y.}~\bibnamefont {Koike}}, \bibinfo {author} {\bibfnamefont
  {M.}~\bibnamefont {Nohara}}, \bibinfo {author} {\bibfnamefont
  {H.}~\bibnamefont {Takagi}}, \bibinfo {author} {\bibfnamefont {Cyril}\
  \bibnamefont {Proust}}, \ and\ \bibinfo {author} {\bibfnamefont {N.~E.}\
  \bibnamefont {Hussey}},\ }\bibfield  {title} {\enquote {\bibinfo {title}
  {{Anomalous Criticality in the Electrical Resistivity of
  La$_{2-x}$Sr$_x$CuO$_4$}},}\ }\href {\doibase 10.1126/science.1165015}
  {\bibfield  {journal} {\bibinfo  {journal} {Science}\ }\textbf {\bibinfo
  {volume} {323}},\ \bibinfo {pages} {603} (\bibinfo {year}
  {2009})}\BibitemShut {NoStop}%
\bibitem [{\citenamefont {Nakatsuji}\ \emph {et~al.}(2003)\citenamefont
  {Nakatsuji}, \citenamefont {Hall}, \citenamefont {Balicas}, \citenamefont
  {Fisk}, \citenamefont {Sugahara}, \citenamefont {Yoshioka},\ and\
  \citenamefont {Maeno}}]{NakatsujiPRL2003}%
  \BibitemOpen
  \bibfield  {author} {\bibinfo {author} {\bibfnamefont {S.}~\bibnamefont
  {Nakatsuji}}, \bibinfo {author} {\bibfnamefont {D.}~\bibnamefont {Hall}},
  \bibinfo {author} {\bibfnamefont {L.}~\bibnamefont {Balicas}}, \bibinfo
  {author} {\bibfnamefont {Z.}~\bibnamefont {Fisk}}, \bibinfo {author}
  {\bibfnamefont {K.}~\bibnamefont {Sugahara}}, \bibinfo {author}
  {\bibfnamefont {M.}~\bibnamefont {Yoshioka}}, \ and\ \bibinfo {author}
  {\bibfnamefont {Y.}~\bibnamefont {Maeno}},\ }\bibfield  {title} {\enquote
  {\bibinfo {title} {{Heavy-Mass Fermi Liquid near a Ferromagnetic Instability
  in Layered Ruthenates}},}\ }\href {\doibase 10.1103/PhysRevLett.90.137202}
  {\bibfield  {journal} {\bibinfo  {journal} {Phys. Rev. Lett.}\ }\textbf
  {\bibinfo {volume} {90}},\ \bibinfo {pages} {137202} (\bibinfo {year}
  {2003})}\BibitemShut {NoStop}%
\bibitem [{\citenamefont {{S. Sugano, Y. Tanabe, and H.
  Kamimura}}(1970)}]{Sugano1970}%
  \BibitemOpen
  \bibfield  {author} {\bibinfo {author} {\bibnamefont {{S. Sugano, Y. Tanabe,
  and H. Kamimura}}},\ }\href@noop {} {\emph {\bibinfo {title} {{Multiplets of
  transition-metal ions in crystals, Pure and Applied Physics}}}},\
  Vol.~\bibinfo {volume} {33}\ (\bibinfo  {publisher} {Academic Press},\
  \bibinfo {address} {New York},\ \bibinfo {year} {1970})\BibitemShut {NoStop}%
\bibitem [{\citenamefont {Zhang}\ and\ \citenamefont
  {Imada}(2007)}]{ZhangPRB2007}%
  \BibitemOpen
  \bibfield  {author} {\bibinfo {author} {\bibfnamefont {Y.~Z.}\ \bibnamefont
  {Zhang}}\ and\ \bibinfo {author} {\bibfnamefont {Masatoshi}\ \bibnamefont
  {Imada}},\ }\bibfield  {title} {\enquote {\bibinfo {title} {{Pseudogap and
  Mott transition studied by cellular dynamical mean-field theory}},}\ }\href
  {\doibase 10.1103/PhysRevB.76.045108} {\bibfield  {journal} {\bibinfo
  {journal} {Phys. Rev. B}\ }\textbf {\bibinfo {volume} {76}},\ \bibinfo
  {pages} {045108} (\bibinfo {year} {2007})}\BibitemShut {NoStop}%
\bibitem [{\citenamefont {Song}\ \emph {et~al.}(2017)\citenamefont {Song},
  \citenamefont {Jiang}, \citenamefont {Lin},\ and\ \citenamefont
  {Zhang}}]{SongPRB2017}%
  \BibitemOpen
  \bibfield  {author} {\bibinfo {author} {\bibfnamefont {Ze-Yi}\ \bibnamefont
  {Song}}, \bibinfo {author} {\bibfnamefont {Xiu-Cai}\ \bibnamefont {Jiang}},
  \bibinfo {author} {\bibfnamefont {Hai-Qing}\ \bibnamefont {Lin}}, \ and\
  \bibinfo {author} {\bibfnamefont {Yu-Zhong}\ \bibnamefont {Zhang}},\
  }\bibfield  {title} {\enquote {\bibinfo {title} {{Distinct nature of
  orbital-selective Mott phases dominated by low-energy local spin
  fluctuations}},}\ }\href {\doibase 10.1103/PhysRevB.96.235119} {\bibfield
  {journal} {\bibinfo  {journal} {Phys. Rev. B}\ }\textbf {\bibinfo {volume}
  {96}},\ \bibinfo {pages} {235119} (\bibinfo {year} {2017})}\BibitemShut
  {NoStop}%
\bibitem [{\citenamefont {Kugler}\ \emph {et~al.}(2019)\citenamefont {Kugler},
  \citenamefont {Lee}, \citenamefont {Weichselbaum}, \citenamefont {Kotliar},\
  and\ \citenamefont {von Delft}}]{KuglerPRB2019}%
  \BibitemOpen
  \bibfield  {author} {\bibinfo {author} {\bibfnamefont {Fabian~B.}\
  \bibnamefont {Kugler}}, \bibinfo {author} {\bibfnamefont {Seung-Sup~B.}\
  \bibnamefont {Lee}}, \bibinfo {author} {\bibfnamefont {Andreas}\ \bibnamefont
  {Weichselbaum}}, \bibinfo {author} {\bibfnamefont {Gabriel}\ \bibnamefont
  {Kotliar}}, \ and\ \bibinfo {author} {\bibfnamefont {Jan}\ \bibnamefont {von
  Delft}},\ }\bibfield  {title} {\enquote {\bibinfo {title} {{Orbital
  differentiation in Hund metals}},}\ }\href {\doibase
  10.1103/PhysRevB.100.115159} {\bibfield  {journal} {\bibinfo  {journal}
  {Phys. Rev. B}\ }\textbf {\bibinfo {volume} {100}},\ \bibinfo {pages}
  {115159} (\bibinfo {year} {2019})}\BibitemShut {NoStop}%
\bibitem [{\citenamefont {Park}\ \emph {et~al.}(2008)\citenamefont {Park},
  \citenamefont {Haule},\ and\ \citenamefont {Kotliar}}]{ParkPRL2008}%
  \BibitemOpen
  \bibfield  {author} {\bibinfo {author} {\bibfnamefont {H.}~\bibnamefont
  {Park}}, \bibinfo {author} {\bibfnamefont {K.}~\bibnamefont {Haule}}, \ and\
  \bibinfo {author} {\bibfnamefont {G.}~\bibnamefont {Kotliar}},\ }\bibfield
  {title} {\enquote {\bibinfo {title} {{Cluster Dynamical Mean Field Theory of
  the Mott Transition}},}\ }\href {\doibase 10.1103/PhysRevLett.101.186403}
  {\bibfield  {journal} {\bibinfo  {journal} {Phys. Rev. Lett.}\ }\textbf
  {\bibinfo {volume} {101}},\ \bibinfo {pages} {186403} (\bibinfo {year}
  {2008})}\BibitemShut {NoStop}%
\bibitem [{\citenamefont {Veenstra}\ \emph {et~al.}(2014)\citenamefont
  {Veenstra}, \citenamefont {Zhu}, \citenamefont {Raichle}, \citenamefont
  {Ludbrook}, \citenamefont {Nicolaou}, \citenamefont {Slomski}, \citenamefont
  {Landolt}, \citenamefont {Kittaka}, \citenamefont {Maeno}, \citenamefont
  {Dil}, \citenamefont {Elfimov}, \citenamefont {Haverkort},\ and\
  \citenamefont {Damascelli}}]{VeenstraPRL2014}%
  \BibitemOpen
  \bibfield  {author} {\bibinfo {author} {\bibfnamefont {C.~N.}\ \bibnamefont
  {Veenstra}}, \bibinfo {author} {\bibfnamefont {Z.-H.}\ \bibnamefont {Zhu}},
  \bibinfo {author} {\bibfnamefont {M.}~\bibnamefont {Raichle}}, \bibinfo
  {author} {\bibfnamefont {B.~M.}\ \bibnamefont {Ludbrook}}, \bibinfo {author}
  {\bibfnamefont {A.}~\bibnamefont {Nicolaou}}, \bibinfo {author}
  {\bibfnamefont {B.}~\bibnamefont {Slomski}}, \bibinfo {author} {\bibfnamefont
  {G.}~\bibnamefont {Landolt}}, \bibinfo {author} {\bibfnamefont
  {S.}~\bibnamefont {Kittaka}}, \bibinfo {author} {\bibfnamefont
  {Y.}~\bibnamefont {Maeno}}, \bibinfo {author} {\bibfnamefont {J.~H.}\
  \bibnamefont {Dil}}, \bibinfo {author} {\bibfnamefont {I.~S.}\ \bibnamefont
  {Elfimov}}, \bibinfo {author} {\bibfnamefont {M.~W.}\ \bibnamefont
  {Haverkort}}, \ and\ \bibinfo {author} {\bibfnamefont {A.}~\bibnamefont
  {Damascelli}},\ }\bibfield  {title} {\enquote {\bibinfo {title}
  {{Spin-Orbital Entanglement and the Breakdown of Singlets and Triplets in
  ${\mathrm{Sr}}_{2}{\mathrm{RuO}}_{4}$ Revealed by Spin- and Angle-Resolved
  Photoemission Spectroscopy}},}\ }\href {\doibase
  10.1103/PhysRevLett.112.127002} {\bibfield  {journal} {\bibinfo  {journal}
  {Phys. Rev. Lett.}\ }\textbf {\bibinfo {volume} {112}},\ \bibinfo {pages}
  {127002} (\bibinfo {year} {2014})}\BibitemShut {NoStop}%
\bibitem [{\citenamefont {Fatuzzo}\ \emph {et~al.}(2015)\citenamefont
  {Fatuzzo}, \citenamefont {Dantz}, \citenamefont {Fatale}, \citenamefont
  {Olalde-Velasco}, \citenamefont {Shaik}, \citenamefont {Dalla~Piazza},
  \citenamefont {Toth}, \citenamefont {Pelliciari}, \citenamefont {Fittipaldi},
  \citenamefont {Vecchione}, \citenamefont {Kikugawa}, \citenamefont {Brooks},
  \citenamefont {R\o{}nnow}, \citenamefont {Grioni}, \citenamefont {R\"uegg},
  \citenamefont {Schmitt},\ and\ \citenamefont {Chang}}]{FatuzzoPRB2015}%
  \BibitemOpen
  \bibfield  {author} {\bibinfo {author} {\bibfnamefont {C.~G.}\ \bibnamefont
  {Fatuzzo}}, \bibinfo {author} {\bibfnamefont {M.}~\bibnamefont {Dantz}},
  \bibinfo {author} {\bibfnamefont {S.}~\bibnamefont {Fatale}}, \bibinfo
  {author} {\bibfnamefont {P.}~\bibnamefont {Olalde-Velasco}}, \bibinfo
  {author} {\bibfnamefont {N.~E.}\ \bibnamefont {Shaik}}, \bibinfo {author}
  {\bibfnamefont {B.}~\bibnamefont {Dalla~Piazza}}, \bibinfo {author}
  {\bibfnamefont {S.}~\bibnamefont {Toth}}, \bibinfo {author} {\bibfnamefont
  {J.}~\bibnamefont {Pelliciari}}, \bibinfo {author} {\bibfnamefont
  {R.}~\bibnamefont {Fittipaldi}}, \bibinfo {author} {\bibfnamefont
  {A.}~\bibnamefont {Vecchione}}, \bibinfo {author} {\bibfnamefont
  {N.}~\bibnamefont {Kikugawa}}, \bibinfo {author} {\bibfnamefont {J.~S.}\
  \bibnamefont {Brooks}}, \bibinfo {author} {\bibfnamefont {H.~M.}\
  \bibnamefont {R\o{}nnow}}, \bibinfo {author} {\bibfnamefont {M.}~\bibnamefont
  {Grioni}}, \bibinfo {author} {\bibfnamefont {Ch.}\ \bibnamefont {R\"uegg}},
  \bibinfo {author} {\bibfnamefont {T.}~\bibnamefont {Schmitt}}, \ and\
  \bibinfo {author} {\bibfnamefont {J.}~\bibnamefont {Chang}},\ }\bibfield
  {title} {\enquote {\bibinfo {title} {{Spin-orbit-induced orbital excitations
  in ${\text{Sr}}_{2}{\text{RuO}}_{4}$ and ${\text{Ca}}_{2}{\text{RuO}}_{4}$: A
  resonant inelastic x-ray scattering study}},}\ }\href {\doibase
  10.1103/PhysRevB.91.155104} {\bibfield  {journal} {\bibinfo  {journal} {Phys.
  Rev. B}\ }\textbf {\bibinfo {volume} {91}},\ \bibinfo {pages} {155104}
  (\bibinfo {year} {2015})}\BibitemShut {NoStop}%
\bibitem [{\citenamefont {Gretarsson}\ \emph {et~al.}(2019)\citenamefont
  {Gretarsson}, \citenamefont {Suzuki}, \citenamefont {Kim}, \citenamefont
  {Ueda}, \citenamefont {Krautloher}, \citenamefont {Kim}, \citenamefont
  {Yava\ifmmode~\mbox{\c{s}}\else \c{s}\fi{}}, \citenamefont {Khaliullin},\
  and\ \citenamefont {Keimer}}]{GretarssonPRB2019}%
  \BibitemOpen
  \bibfield  {author} {\bibinfo {author} {\bibfnamefont {H.}~\bibnamefont
  {Gretarsson}}, \bibinfo {author} {\bibfnamefont {H.}~\bibnamefont {Suzuki}},
  \bibinfo {author} {\bibfnamefont {Hoon}\ \bibnamefont {Kim}}, \bibinfo
  {author} {\bibfnamefont {K.}~\bibnamefont {Ueda}}, \bibinfo {author}
  {\bibfnamefont {M.}~\bibnamefont {Krautloher}}, \bibinfo {author}
  {\bibfnamefont {B.~J.}\ \bibnamefont {Kim}}, \bibinfo {author} {\bibfnamefont
  {H.}~\bibnamefont {Yava\ifmmode~\mbox{\c{s}}\else \c{s}\fi{}}}, \bibinfo
  {author} {\bibfnamefont {G.}~\bibnamefont {Khaliullin}}, \ and\ \bibinfo
  {author} {\bibfnamefont {B.}~\bibnamefont {Keimer}},\ }\bibfield  {title}
  {\enquote {\bibinfo {title} {{Observation of spin-orbit excitations and
  Hund's multiplets in ${\mathrm{Ca}}_{2}{\mathrm{RuO}}_{4}$}},}\ }\href
  {\doibase 10.1103/PhysRevB.100.045123} {\bibfield  {journal} {\bibinfo
  {journal} {Phys. Rev. B}\ }\textbf {\bibinfo {volume} {100}},\ \bibinfo
  {pages} {045123} (\bibinfo {year} {2019})}\BibitemShut {NoStop}%
\bibitem [{\citenamefont {Friedt}\ \emph {et~al.}(2001)\citenamefont {Friedt},
  \citenamefont {Braden}, \citenamefont {Andr\'e}, \citenamefont {Adelmann},
  \citenamefont {Nakatsuji},\ and\ \citenamefont {Maeno}}]{FriedtPRB2001}%
  \BibitemOpen
  \bibfield  {author} {\bibinfo {author} {\bibfnamefont {O.}~\bibnamefont
  {Friedt}}, \bibinfo {author} {\bibfnamefont {M.}~\bibnamefont {Braden}},
  \bibinfo {author} {\bibfnamefont {G.}~\bibnamefont {Andr\'e}}, \bibinfo
  {author} {\bibfnamefont {P.}~\bibnamefont {Adelmann}}, \bibinfo {author}
  {\bibfnamefont {S.}~\bibnamefont {Nakatsuji}}, \ and\ \bibinfo {author}
  {\bibfnamefont {Y.}~\bibnamefont {Maeno}},\ }\bibfield  {title} {\enquote
  {\bibinfo {title} {{Structural and magnetic aspects of the metal-insulator
  transition in
  ${\mathrm{Ca}}_{2\ensuremath{-}x}{\mathrm{Sr}}_{x}{\mathrm{RuO}}_{4}$}},}\
  }\href {\doibase 10.1103/PhysRevB.63.174432} {\bibfield  {journal} {\bibinfo
  {journal} {Phys. Rev. B}\ }\textbf {\bibinfo {volume} {63}},\ \bibinfo
  {pages} {174432} (\bibinfo {year} {2001})}\BibitemShut {NoStop}%
\bibitem [{\citenamefont {Kim}\ \emph {et~al.}(2018)\citenamefont {Kim},
  \citenamefont {Mravlje}, \citenamefont {Ferrero}, \citenamefont {Parcollet},\
  and\ \citenamefont {Georges}}]{KimPRL2018}%
  \BibitemOpen
  \bibfield  {author} {\bibinfo {author} {\bibfnamefont {Minjae}\ \bibnamefont
  {Kim}}, \bibinfo {author} {\bibfnamefont {Jernej}\ \bibnamefont {Mravlje}},
  \bibinfo {author} {\bibfnamefont {Michel}\ \bibnamefont {Ferrero}}, \bibinfo
  {author} {\bibfnamefont {Olivier}\ \bibnamefont {Parcollet}}, \ and\ \bibinfo
  {author} {\bibfnamefont {Antoine}\ \bibnamefont {Georges}},\ }\bibfield
  {title} {\enquote {\bibinfo {title} {{Spin-Orbit Coupling and Electronic
  Correlations in ${\mathrm{Sr}}_{2}{\mathrm{RuO}}_{4}$}},}\ }\href {\doibase
  10.1103/PhysRevLett.120.126401} {\bibfield  {journal} {\bibinfo  {journal}
  {Phys. Rev. Lett.}\ }\textbf {\bibinfo {volume} {120}},\ \bibinfo {pages}
  {126401} (\bibinfo {year} {2018})}\BibitemShut {NoStop}%
\bibitem [{\citenamefont {Zhang}\ and\ \citenamefont
  {Pavarini}(2017)}]{ZhangPRB2017}%
  \BibitemOpen
  \bibfield  {author} {\bibinfo {author} {\bibfnamefont {Guoren}\ \bibnamefont
  {Zhang}}\ and\ \bibinfo {author} {\bibfnamefont {Eva}\ \bibnamefont
  {Pavarini}},\ }\bibfield  {title} {\enquote {\bibinfo {title} {{Mott
  transition, spin-orbit effects, and magnetism in
  ${\mathrm{Ca}}_{2}{\mathrm{RuO}}_{4}$}},}\ }\href {\doibase
  10.1103/PhysRevB.95.075145} {\bibfield  {journal} {\bibinfo  {journal} {Phys.
  Rev. B}\ }\textbf {\bibinfo {volume} {95}},\ \bibinfo {pages} {075145}
  (\bibinfo {year} {2017})}\BibitemShut {NoStop}%
\bibitem [{\citenamefont {Burganov}\ \emph {et~al.}(2016)\citenamefont
  {Burganov}, \citenamefont {Adamo}, \citenamefont {Mulder}, \citenamefont
  {Uchida}, \citenamefont {King}, \citenamefont {Harter}, \citenamefont {Shai},
  \citenamefont {Gibbs}, \citenamefont {Mackenzie}, \citenamefont {Uecker},
  \citenamefont {Bruetzam}, \citenamefont {Beasley}, \citenamefont {Fennie},
  \citenamefont {Schlom},\ and\ \citenamefont {Shen}}]{BurganovPRL2016}%
  \BibitemOpen
  \bibfield  {author} {\bibinfo {author} {\bibfnamefont {B.}~\bibnamefont
  {Burganov}}, \bibinfo {author} {\bibfnamefont {C.}~\bibnamefont {Adamo}},
  \bibinfo {author} {\bibfnamefont {A.}~\bibnamefont {Mulder}}, \bibinfo
  {author} {\bibfnamefont {M.}~\bibnamefont {Uchida}}, \bibinfo {author}
  {\bibfnamefont {P.~D.~C.}\ \bibnamefont {King}}, \bibinfo {author}
  {\bibfnamefont {J.~W.}\ \bibnamefont {Harter}}, \bibinfo {author}
  {\bibfnamefont {D.~E.}\ \bibnamefont {Shai}}, \bibinfo {author}
  {\bibfnamefont {A.~S.}\ \bibnamefont {Gibbs}}, \bibinfo {author}
  {\bibfnamefont {A.~P.}\ \bibnamefont {Mackenzie}}, \bibinfo {author}
  {\bibfnamefont {R.}~\bibnamefont {Uecker}}, \bibinfo {author} {\bibfnamefont
  {M.}~\bibnamefont {Bruetzam}}, \bibinfo {author} {\bibfnamefont {M.~R.}\
  \bibnamefont {Beasley}}, \bibinfo {author} {\bibfnamefont {C.~J.}\
  \bibnamefont {Fennie}}, \bibinfo {author} {\bibfnamefont {D.~G.}\
  \bibnamefont {Schlom}}, \ and\ \bibinfo {author} {\bibfnamefont {K.~M.}\
  \bibnamefont {Shen}},\ }\bibfield  {title} {\enquote {\bibinfo {title}
  {{Strain Control of Fermiology and Many-Body Interactions in Two-Dimensional
  Ruthenates}},}\ }\href {\doibase 10.1103/PhysRevLett.116.197003} {\bibfield
  {journal} {\bibinfo  {journal} {Phys. Rev. Lett.}\ }\textbf {\bibinfo
  {volume} {116}},\ \bibinfo {pages} {197003} (\bibinfo {year}
  {2016})}\BibitemShut {NoStop}%
\bibitem [{\citenamefont {Barber}\ \emph {et~al.}(2018)\citenamefont {Barber},
  \citenamefont {Gibbs}, \citenamefont {Maeno}, \citenamefont {Mackenzie},\
  and\ \citenamefont {Hicks}}]{BarberPRL2018}%
  \BibitemOpen
  \bibfield  {author} {\bibinfo {author} {\bibfnamefont {M.~E.}\ \bibnamefont
  {Barber}}, \bibinfo {author} {\bibfnamefont {A.~S.}\ \bibnamefont {Gibbs}},
  \bibinfo {author} {\bibfnamefont {Y.}~\bibnamefont {Maeno}}, \bibinfo
  {author} {\bibfnamefont {A.~P.}\ \bibnamefont {Mackenzie}}, \ and\ \bibinfo
  {author} {\bibfnamefont {C.~W.}\ \bibnamefont {Hicks}},\ }\bibfield  {title}
  {\enquote {\bibinfo {title} {{Resistivity in the Vicinity of a van Hove
  Singularity: ${\mathrm{Sr}}_{2}{\mathrm{RuO}}_{4}$ under Uniaxial
  Pressure}},}\ }\href {\doibase 10.1103/PhysRevLett.120.076602} {\bibfield
  {journal} {\bibinfo  {journal} {Phys. Rev. Lett.}\ }\textbf {\bibinfo
  {volume} {120}},\ \bibinfo {pages} {076602} (\bibinfo {year}
  {2018})}\BibitemShut {NoStop}%
\bibitem [{\citenamefont {Lee}\ \emph {et~al.}(2002)\citenamefont {Lee},
  \citenamefont {Yu}, \citenamefont {Lee}, \citenamefont {Noh}, \citenamefont
  {Gimm}, \citenamefont {Choi},\ and\ \citenamefont {Eom}}]{LeePRB2002}%
  \BibitemOpen
  \bibfield  {author} {\bibinfo {author} {\bibfnamefont {Y.~S.}\ \bibnamefont
  {Lee}}, \bibinfo {author} {\bibfnamefont {Jaejun}\ \bibnamefont {Yu}},
  \bibinfo {author} {\bibfnamefont {J.~S.}\ \bibnamefont {Lee}}, \bibinfo
  {author} {\bibfnamefont {T.~W.}\ \bibnamefont {Noh}}, \bibinfo {author}
  {\bibfnamefont {T.-H.}\ \bibnamefont {Gimm}}, \bibinfo {author}
  {\bibfnamefont {Han-Yong}\ \bibnamefont {Choi}}, \ and\ \bibinfo {author}
  {\bibfnamefont {C.~B.}\ \bibnamefont {Eom}},\ }\bibfield  {title} {\enquote
  {\bibinfo {title} {{Non-Fermi liquid behavior and scaling of the
  low-frequency suppression in the optical conductivity spectra of
  ${\mathrm{CaRuO}}_{3}$}},}\ }\href {\doibase 10.1103/PhysRevB.66.041104}
  {\bibfield  {journal} {\bibinfo  {journal} {Phys. Rev. B}\ }\textbf {\bibinfo
  {volume} {66}},\ \bibinfo {pages} {041104} (\bibinfo {year}
  {2002})}\BibitemShut {NoStop}%
\bibitem [{\citenamefont {Kamal}\ \emph {et~al.}(2006)\citenamefont {Kamal},
  \citenamefont {Kim}, \citenamefont {Eom},\ and\ \citenamefont
  {Dodge}}]{KamalPRB2006}%
  \BibitemOpen
  \bibfield  {author} {\bibinfo {author} {\bibfnamefont {Saeid}\ \bibnamefont
  {Kamal}}, \bibinfo {author} {\bibfnamefont {D.~M.}\ \bibnamefont {Kim}},
  \bibinfo {author} {\bibfnamefont {C.~B.}\ \bibnamefont {Eom}}, \ and\
  \bibinfo {author} {\bibfnamefont {J.~S.}\ \bibnamefont {Dodge}},\ }\bibfield
  {title} {\enquote {\bibinfo {title} {{Terahertz-frequency carrier dynamics
  and spectral weight redistribution in the nearly magnetic metal
  $\mathrm{Ca}\mathrm{Ru}{\mathrm{O}}_{3}$}},}\ }\href {\doibase
  10.1103/PhysRevB.74.165115} {\bibfield  {journal} {\bibinfo  {journal} {Phys.
  Rev. B}\ }\textbf {\bibinfo {volume} {74}},\ \bibinfo {pages} {165115}
  (\bibinfo {year} {2006})}\BibitemShut {NoStop}%
\bibitem [{\citenamefont {Schneider}\ \emph {et~al.}(2014)\citenamefont
  {Schneider}, \citenamefont {Geiger}, \citenamefont {Esser}, \citenamefont
  {Pracht}, \citenamefont {Stingl}, \citenamefont {Tokiwa}, \citenamefont
  {Moshnyaga}, \citenamefont {Sheikin}, \citenamefont {Mravlje}, \citenamefont
  {Scheffler},\ and\ \citenamefont {Gegenwart}}]{SchneiderPRL2014}%
  \BibitemOpen
  \bibfield  {author} {\bibinfo {author} {\bibfnamefont {M.}~\bibnamefont
  {Schneider}}, \bibinfo {author} {\bibfnamefont {D.}~\bibnamefont {Geiger}},
  \bibinfo {author} {\bibfnamefont {S.}~\bibnamefont {Esser}}, \bibinfo
  {author} {\bibfnamefont {U.~S.}\ \bibnamefont {Pracht}}, \bibinfo {author}
  {\bibfnamefont {C.}~\bibnamefont {Stingl}}, \bibinfo {author} {\bibfnamefont
  {Y.}~\bibnamefont {Tokiwa}}, \bibinfo {author} {\bibfnamefont
  {V.}~\bibnamefont {Moshnyaga}}, \bibinfo {author} {\bibfnamefont
  {I.}~\bibnamefont {Sheikin}}, \bibinfo {author} {\bibfnamefont
  {J.}~\bibnamefont {Mravlje}}, \bibinfo {author} {\bibfnamefont
  {M.}~\bibnamefont {Scheffler}}, \ and\ \bibinfo {author} {\bibfnamefont
  {P.}~\bibnamefont {Gegenwart}},\ }\bibfield  {title} {\enquote {\bibinfo
  {title} {{Low-Energy Electronic Properties of Clean ${\mathrm{CaRuO}}_{3}$:
  Elusive Landau Quasiparticles}},}\ }\href {\doibase
  10.1103/PhysRevLett.112.206403} {\bibfield  {journal} {\bibinfo  {journal}
  {Phys. Rev. Lett.}\ }\textbf {\bibinfo {volume} {112}},\ \bibinfo {pages}
  {206403} (\bibinfo {year} {2014})}\BibitemShut {NoStop}%
\bibitem [{\citenamefont {Yang}\ \emph {et~al.}(2016)\citenamefont {Yang},
  \citenamefont {Fan}, \citenamefont {Liu}, \citenamefont {Yao}, \citenamefont
  {Li}, \citenamefont {Liu}, \citenamefont {Jiang},\ and\ \citenamefont
  {Shen}}]{YangPRB2016}%
  \BibitemOpen
  \bibfield  {author} {\bibinfo {author} {\bibfnamefont {H.~F.}\ \bibnamefont
  {Yang}}, \bibinfo {author} {\bibfnamefont {C.~C.}\ \bibnamefont {Fan}},
  \bibinfo {author} {\bibfnamefont {Z.~T.}\ \bibnamefont {Liu}}, \bibinfo
  {author} {\bibfnamefont {Q.}~\bibnamefont {Yao}}, \bibinfo {author}
  {\bibfnamefont {M.~Y.}\ \bibnamefont {Li}}, \bibinfo {author} {\bibfnamefont
  {J.~S.}\ \bibnamefont {Liu}}, \bibinfo {author} {\bibfnamefont {M.~H.}\
  \bibnamefont {Jiang}}, \ and\ \bibinfo {author} {\bibfnamefont {D.~W.}\
  \bibnamefont {Shen}},\ }\bibfield  {title} {\enquote {\bibinfo {title}
  {{Comparative angle-resolved photoemission spectroscopy study of
  ${\mathrm{CaRuO}}_{3}$ and ${\mathrm{SrRuO}}_{3}$ thin films: Pronounced
  spectral weight transfer and possible precursor of lower Hubbard band}},}\
  }\href {\doibase 10.1103/PhysRevB.94.115151} {\bibfield  {journal} {\bibinfo
  {journal} {Phys. Rev. B}\ }\textbf {\bibinfo {volume} {94}},\ \bibinfo
  {pages} {115151} (\bibinfo {year} {2016})}\BibitemShut {NoStop}%
\bibitem [{\citenamefont {Liu}\ \emph {et~al.}(2018)\citenamefont {Liu},
  \citenamefont {Nair}, \citenamefont {Ruf}, \citenamefont {Schlom},\ and\
  \citenamefont {Shen}}]{LiuPRB2018}%
  \BibitemOpen
  \bibfield  {author} {\bibinfo {author} {\bibfnamefont {Yang}\ \bibnamefont
  {Liu}}, \bibinfo {author} {\bibfnamefont {Hari~P.}\ \bibnamefont {Nair}},
  \bibinfo {author} {\bibfnamefont {Jacob~P.}\ \bibnamefont {Ruf}}, \bibinfo
  {author} {\bibfnamefont {Darrell~G.}\ \bibnamefont {Schlom}}, \ and\ \bibinfo
  {author} {\bibfnamefont {Kyle~M.}\ \bibnamefont {Shen}},\ }\bibfield  {title}
  {\enquote {\bibinfo {title} {{Revealing the hidden heavy Fermi liquid in
  ${\mathrm{CaRuO}}_{3}$}},}\ }\href {\doibase 10.1103/PhysRevB.98.041110}
  {\bibfield  {journal} {\bibinfo  {journal} {Phys. Rev. B}\ }\textbf {\bibinfo
  {volume} {98}},\ \bibinfo {pages} {041110} (\bibinfo {year}
  {2018})}\BibitemShut {NoStop}%
\end{thebibliography}%
\end{document}